\documentclass[a4paper,11pt]{article}
\pdfoutput=1 

\usepackage{jcappub} 

\usepackage[T1]{fontenc} 
\usepackage{journals}
\usepackage{bm}
\usepackage{multirow}
\def \mpc {\mbox{\rm Mpc}}

\title{\boldmath CLASH-VLT: constraints on  $f(R)$ gravity models with galaxy clusters using lensing and kinematic analyses}

\author[1]{L. Pizzuti,}
\author[1,2]{B. Sartoris,}
\author[4]{ L. Amendola,}
\author[1,2,3]{S. Borgani,}
\author[2]{A. Biviano,}
\author[5]{K. Umetsu,}
\author[6]{A. Mercurio,}
\author[7]{P. Rosati,}

\author[2]{I. Balestra,}
\author[7]{G. B. Caminha,}
\author[1,2]{M. Girardi,}
\author[7,8]{C. Grillo,}
\author[2]{M. Nonino.}

\affiliation[1]{Dipartimento di Fisica, Sezione di Astronomia, Universit\`a di Trieste,\\ Via Tiepolo 11, I-34143 Trieste, Italy}
\affiliation[2]{INAF - Osservatorio Astronomico di Trieste,\\ Via Tiepolo 11, I-34143 Trieste, Italy}
\affiliation[3]{ INFN - Sezione di Trieste,\\ Via Valerio 2, I-34127 Trieste, Italy}
\affiliation[4]{Institut f\"ur Theoretische Physik, Universit\"at Heidelberg,\\ Philosophenweg 16, D-69120 Heidelberg, Germany}
\affiliation[5]{5 Institute of Astronomy and Astrophysics, Academia Sinica,
P.O. Box 23-141,\\ Taipei 10617, Taiwan}
\affiliation[6]{Osservatorio Astronomico di Capodimonte, \\Via Moiariello 16, I-80131 Napoli, Italy}
\affiliation[7]{Dipartimento di Fisica, Universit\`a degli Studi di Milano, via Celoria 16, I-20133 Milano, Italy}
\affiliation[8]{ Dark Cosmology Centre, Niels Bohr Institute, University of Copenhagen,\\ Juliane Maries Vej 30, DK-2100 Copenhagen, Denmark}

\emailAdd{pizzuti@oats.inaf.it}
\emailAdd{sartoris@oats.inaf.it }
\emailAdd{l.amendola@thphys.uni-heidelberg.de}
\emailAdd{borgani@oats.inaf.it}
\emailAdd{biviano@oats.inaf.it}
\emailAdd{keiichi@asiaa.sinica.edu.tw}
\emailAdd{mercurio@na.astro.it}
\emailAdd{rosati@fe.infn.it}
\emailAdd{balestra@oats.inaf.it}
\emailAdd{gbcaminha@gmail.com}
\emailAdd{girardi@oats.inaf.it}
\emailAdd{grillo.claudio@googlemail.com}
\emailAdd{nonino@oats.inaf.it}

\abstract{We perform a maximum likelihood kinematic analysis of the two dynamically
relaxed galaxy clusters MACS J1206.2-0847 at $z=0.44$ and RXC J2248.7-4431
at $z=0.35$ to determine the total mass profile in modified gravity
models, using a modified version of the MAMPOSSt code of Mamon, Biviano
and Bou\'e. Our work is based on the kinematic and
lensing mass profiles derived using the data from the Cluster Lensing
And Supernova survey with Hubble (hereafter CLASH) and the spectroscopic
follow-up with the Very Large Telescope (hereafter CLASH-VLT). We assume a spherical Navarro-Frenk-White (NFW hereafter)
profile in order to obtain a constraint on the fifth force interaction
range $\lambda$ for models in which the dependence of this parameter on the environment
is negligible at the scale considered (i.e. $\lambda=const$) and fixing
the fifth force strength to the value predicted in $f(R)$ gravity.
We then use information from lensing analysis to put a prior on the
other NFW free parameters. In
the case of MACSJ 1206 the joint kinematic+lensing analysis leads
to an upper limit on the effective interaction range $\lambda\le1.61\,\mpc$
at $\Delta\chi^{2}=2.71$ on the marginalized distribution. For RXJ
2248 instead a possible tension with the $\Lambda$CDM model appears
when adding lensing information, with a lower limit $\lambda\ge0.14\,\mpc$
at $\Delta\chi^{2}=2.71$. This is consequence of the slight difference
between the lensing and kinematic data, appearing in GR for this cluster, that could in principle be explained in terms of modifications of gravity. We discuss the
impact of systematics and the limits of our analysis as well as future
improvements of the results obtained. This work has interesting implications in view of upcoming and future large imaging and spectroscopic surveys, that will deliver lensing and kinematic mass reconstruction for a large number of galaxy clusters.}

\begin{document}
\maketitle
\flushbottom
\section{Introduction}

The discovery of the late-time accelerated expansion of the Universe
 (refs. \cite{Reiss01,Perlmutter99}) at the turn of the XX century
has posed one of the most challenging problems in modern physics.
The current cosmological concordance model, or $\Lambda$CDM model,
correctly predicts the observed expansion history but requires the
introduction of a cosmological constant $\Lambda$, an additional
term in the Einstein equations that has no natural explanation in
the framework of standard physics. In the last decade several alternatives
have been proposed to explain the origin of the acceleration; a possible
solution is to modify the theory of General Relativity (GR hereafter)
on large scales by introducing new degrees of freedom which can reproduce
the same effect of a cosmological constant (see e.g. refs. \cite{Hu2007,Dvali07})
and at the same time propagate a new force. Even when considering
theories where the predicted expansion history is identical to that
of the $\Lambda$CDM model, the evolution of the perturbations
can be significantly different from $\Lambda$CDM.

General departures from GR change the relations between the gravitational
potentials $\Phi$ and $\Psi$ in the linearly perturbed Friedmann-Lemaitre-Robertson-Walker
metric and the fluctuations of the matter density field both at linear
and non-linear level (ref. \cite{Lue01}). We can generally parametrize
these deviations in terms of two dimensionless functions $\eta(k,a)$,
$Y(k,a)$ (see e.g. ref. \cite{Z09}) or their combination, which
can be constrained with a broad range of observational probes, such
as Cosmic Microwave Background anisotropies (refs. \cite{H2013,Za2010,Planckmod}),
Barionic Acoustic Oscillations (e.g. ref. \cite{Ya2006}), redshift
space distortions (e.g. ref. \cite{J2012,Z09}), galaxy clusters (e.g.
refs. \cite{Ferraro11,Pizzuti16,Cataneo}). In particular, the analysis
of ref. \cite{Planckmod}, further confirmed by ref. \cite{Dival2016},
seems to indicate a tension in the amount of clustering with the concordance
model when CMB data are combined with low-redshift probes (galaxy
weak lensing and redshift space distortions) that could be explained
in terms of modified gravity.

Since GR is tested at high precision at Solar System scales (see e.g.
ref. \cite{Will06} and references therein), modified gravity (hereafter
MG) models must match the standard theory of gravity in this regime.
Solar System constraints on modification of gravity ca be accounted for a screening mechanism
which suppresses the modifications restoring GR in high density environments
(see e.g. ref \cite{K2010} for a review). Alternatively, local gravity
constraints can be escaped by assuming that baryons are decoupled (ref. \cite{Amendola2000}).

Among the modified gravity models, a very large class is represented
by the Horndeski lagrangian (ref. \cite{Horn1974}), the most general
theory of massless gravity and a single scalar field with second-order
equations of motion. In the quasi-static regime, i.e. when the wave
nature of the scalar degree of freedom can be neglected, the extra
force carried by the scalar field manifests itself as a Yukawa correction
to the Newtonian potential, characterized by two parameters, a strength
$Q$ and a range $\lambda$. Aim of this paper is to constrain these
parameters by employing reconstruction of the mass profiles of galaxy clusters
with kinematic analysis and strong+weak lensing analysis. To simplify
our task we apply the method to a sub-class of Horndeski models, the
so-called $f(R)$ models, in which case it turns out that $2Q^{2}=1/3$
and also that the lensing dynamics is not distorted. In
these models the role of the additional degree of freedom is played
by $f_{,R}=df/dR$ that acts as a scalar field with a characteristic
mass $m_{f_{R}}=1/\lambda$. The new force is suppressed in high density
regions by a non-linear mechanism that quenches deviations from GR,
as required by consistency with the current observations, known as
chameleon screening (ref. %
\mbox{%
\cite{K2004}%
}).
 
We aim then at constraining the interaction range $\lambda$ by performing
a joint kinematics and lensing analysis of the galaxy clusters, MACS
J1206.2-0847 (hereafter MACS 1206) at redshift $z=0.44$ and RXC J2248.7-4431
(hereafter RXJ 2248) at $z=0.35$, which have been analysed in detail
as part of the Cluster Lensing And Supernova survey with Hubble (CLASH,
ref. \cite{Postman01}) and the spectroscopic follow-up with the Very
Large Telescope (CLASH-VLT, ref. \cite{Rosati1}) programs. We determine
the dynamic mass profiles in $f(R)$ gravity under the assumptions
of spherical symmetry and dynamical relaxation of the clusters by using the
\emph{MAMPOSSt} code of ref. \cite{Mamon01}, in which we included
a parametric expression of the gravitational potential valid for generic
MG models. This expression is obtained by imposing a NFW profile (ref.
\cite{Navarro}) for the matter density perturbation. 
 With a Maximum Likelihood approach we constrain the free
parameters in our analysis, namely the scale radius $r_{s}$, the
radius $r_{200}$ (a proxy for the virial radius), the parameter describing the velocity anisotropy profile $\beta(r)$ and
the interaction range $\lambda$. We assume a constant value for the
scalaron mass $m_{f_{R}}=1/\lambda$, which means we are neglecting
the change in $m_{f_{R}}$ due to the environmental density. This
can be translated as dealing with models where the screening mechanism
takes place at scales much smaller than the cluster size (e.g. few
kpc), but also with models for which the screening is so effective
that maintains the field mass nearly constant to the value inside
the overdensity. In this case, the results we obtain on $\lambda$ refer to
an effective \textquotedbl{}screened\textquotedbl{} $f_{,R}$, which
is much smaller than the background field (ref. \cite{Li12}). 
We further combine the \emph{MAMPOSSt} Likelihood with the posterior
probability distribution of the NFW parameters $r_{s}$ and $r_{200}$
obtained by the joint strong+weak lensing analysis of ref. \cite{Umetsu16}
in order to improve our results on $\lambda$.

The paper is organized as follows: in Sect. \ref{sec:theory} we discuss
the general parametrization of the metric potentials in modified gravity,
the application on the $f(R)$ class of MG models. 
In Sect \ref{sec:MAM} we present the \emph{MAMPOSSt} method and the modifications we made to the code. Sect. \ref{sec:cluster} is
dedicated to describe the properties of the two analysed galaxy clusters.
In Sect. \ref{sec:results} we show our results, which are further
discussed in Sect. \ref{sec:discussion}, where we also draw our main conclusions. Throughout this paper we
assume a flat $\Lambda$CDM universe with $\Omega_{m}=0.3$ for the
matter density parameter and $H_{0}=70$ km s$^{-1}$Mpc$^{-1}$ for
the present-day Hubble constant to convert observed angular scales
into physical scales.

\section{Theoretical framework}

\label{sec:theory}

\subsection{General parametrization}

\label{sec:gen} The spacetime structure of a galaxy cluster is well
described by a linear perturbation of the Friedmann Lemaitre Robertson
Walker (FLRW hereafter) metric, which in spherical coordinates is
given by: 
\begin{equation}
ds^{2}=a^{2}(\tau)\left\{ \left(1+2\frac{\Phi}{c^{2}}\right)d\tau^{2}-\left(1-2\frac{\Psi}{c^{2}}\right)[d\chi^{2}+f_{K}^{2}(\chi)d\Omega^{2}]\right\} ,\label{eq:NewGauge}
\end{equation}
where we adopted the conformal Newtonian gauge choice (see ref. \cite{Mukhanov01}).
Although we assume $\Phi,\Psi\ll1$, we do not restrict the value
of the matter perturbations. $\Phi$ and $\Psi$ are the scalar Bardeen
potentials defined in ref. \cite{Bardeen01}, $a(\tau)$ is the expansion
factor, which is a function of the conformal time $\tau$. $f_{K}(\chi)$
is a function of the curvature $K$ and the radial comoving coordinate
$\chi$, with $f_{K}(\chi)=\chi$ for a flat background universe with
$K=0$.

In General Relativity
$\Phi=\Psi\equiv\Phi_{N}$,
where $\Phi_{N}$ is the usual Newtonian potential obeying the Poisson
equation (in Fourier space): 
\begin{equation}
k^{2}\Phi_{N}=-\frac{3}{2}\Omega_{m}(t)\delta_{m}H^{2}.\label{eq:Poisson}
\end{equation}
In the expression above $k=k_{com}/a$ is the physical wavenumber,
$\Omega_{m}(t)$ is the time dependent matter density parameter, $H$ is the Hubble
parameter, $\delta_{m}=(\rho_{m}-\rho_{bg})/\rho_{bg}$ the matter
density contrast, where $\rho_{bg}$ is the background matter density at that time. Here we have already assumed spherical symmetry,
i.e. $\Phi_{N}=\Phi_{N}(k),\,\delta_{m}=\delta_{m}(k)$. In a general
modified gravity scenario, $\Phi$ is no longer equal to $\Psi$, and
the Poisson equation should be changed as (see e.g. refs. \cite{Amendola2012},
\cite{Defelice2011}): 
\begin{equation}
k^{2}\Phi=-\frac{3}{2}Y(k,a)\Omega_{m}(t)\delta_{m}H^{2},\label{eq:phi}
\end{equation}
\begin{equation}
k^{2}(\Phi+\Psi)=-\frac{3}{2}Y(k,a)\left[1+\eta(k,a)\right]\Omega_{m}(t)\delta_{m}H^{2}.\label{eq:psi}
\end{equation}
Here $Y(k,a)$ is the effective gravitational constant and $\eta(k,a)=\Psi/\Phi$
is the anisotropic stress, both equal to unity in standard gravity.
As shown in ref. \cite{Amendola2012}, we can write these quantities
in terms of five parameters introduced in ref. \cite{Defelice2011}:
\begin{align}
 & Y(k,a)=h_{1}\left(\frac{1+k^{2}h_{5}}{1+k^{2}h_{3}}\right), & \eta(k,a)=h_{2}\left(\frac{1+k^{2}h_{4}}{1+k^{2}h_{5}}\right).
\end{align}
Here, $h_{1}...h_{5}$ are functions of time only and can be
considered constant for small redshift ranges, as
the one spanned by our clusters. It's useful to define $Q^{2}=(h_{5}-h_{3})/2h_{3}$
and $\hat{Q}^{2}=(h_{4}-h_{3})/2h_{3}$ such that: 
\begin{align}
 & Y\eta=h_{1}h_{2}\left(1+\frac{2\hat{Q}^{2}k^{2}}{m^{2}+k^{2}}\right), & Y=h_{1}\left(1+\frac{2Q^{2}k^{2}}{m^{2}+k^{2}}\right),
\end{align}
which are in the form of Yukawa potentials with strength $\hat{Q}^{2}$
and $Q^{2}$ respectively and characteristic mass $m^{2}=1/h_{3}$.
Scale-independent standard gravity is recovered for $m\to\infty$
and $h_{1},h_{2}\to1$.

Inserting the Navarro-Frenk-White density profile of ref. \cite{navarro97}
in the RHS of \eqref{eq:phi} and taking the anti-Fourier transform,
the expression for the time-time Bardeen potential in real space could
be written as 
\begin{equation}
\Phi(r)=h_{1}\left[G\int_{r_{0}}^{r}{\frac{dx}{x^{2}}M(x)} +2Q^2\phi_{mg}(r)\right].\label{eq:phimod}
\end{equation}
The first term in eq. \eqref{eq:phimod} is the Newtonian potential $\Phi_N$; inside the integral, $M(x)$ is the
standard-gravity NFW mass profile: 
\begin{equation}
M(r)=M_{200}\left[\log(1+r/r_{s})-r/r_{s}(1+r/r_{s})^{-1}\right]\times\left[\log(1+c)-c/(1+c)\right]^{-1},
\end{equation}
described by the scale radius $r_{s}$ and the parameter $r_{200}$
which is the radius of a sphere enclosing an overdensity 200 times
the critical density at the cluster redshift; $M_{200}$ is the mass
at $r_{200}$ and $c=r_{200}/r_{s}$ is the concentration. The
second term includes the contribution of GR departures and it is 
given by: 
\begin{equation*}
\phi_{mg}(r)= \frac{2\pi G\rho_0}{r}r_s^3\left\{e^{-m(r_s+r)}\left[{\rm Ei}(m\,r_s)-{\rm Ei}(m(r_s+r))\right]\right.
\end{equation*}
\begin{equation}\label{eq:modmass}
\left.-e^{m(r_s+r)}{\rm Ei}\left[-m(r_s+r)\right]+e^{m(r_s-r)}{\rm Ei}(-m\,r_s)\right\},
\end{equation}
where $\rho_{0}$ is the normalization of the NFW density profile
and is a function of the parameters $r_{s},\,r_{200}$; $m$ is the
characteristic mass defined above.
As $m\to0$, it can be shown numerically that the additional
contribution $\phi_ {mg}(r)$ tends to be equal to $\Phi_N(r)$ for a given
$r$.

Combining eq. \eqref{eq:phi} and eq. \eqref{eq:psi}, we obtain a
similar expression for the space-space potential $\Psi$: 
\begin{equation}
\Psi(r)=h_{1}h_{2}\left[G\int_{r_{0}}^{r}{\frac{dx}{x^{2}}M(x)} +2\hat{Q}^2\phi_{mg}(r)\right].
\end{equation}
The effect of modifications of gravity is thus completely determined
by the choice of the parameters $h_{1},h_{2},Q,\hat{Q},m$. In the
general case they are totally independent from each other, but, as
we will see in the next paragraph, fixing a particular model allows
us to establish relations among these parameters, reducing the number
of degrees of freedom.

Since in a galaxy cluster the typical velocities of the galaxies are
non-relativistic ($\sim10^{3}\,km/s<<c$), in the weak field limit
the motion of the galaxies is determined only by the time-time component
of the metric. As follows from the geodesics equation, the acceleration
experienced by a non-relativistic particle is sourced by the gradient
of $\Phi$: 
\begin{equation}
\frac{d^{2}\vec{x}}{dt^{2}}=-\nabla\Phi.
\end{equation}
It is thus possible to infer the Bardeen potential $\Phi$ by the dynamical
analysis of the observed galaxies in a cluster. The combination of
$\Psi$ and $\Phi$ can instead be determined through gravitational
lensing observations. Indeed, photons perceive the gravitational potential
due to the contribution of both time-time and space-space metric components.
As in ref. \cite{Pizzuti16}, we define at leading order in $\Psi,\Phi$
a lensing potential $\Phi_{lens}=(\Psi+\Phi)/2$, which is related
to the lensing mass density profile through the Poisson equation 
\begin{equation}
\nabla^{2}\Phi_{lens}=4\pi G\rho_{lens}\,.\label{eq:lens}
\end{equation}
Hence, gravitational lensing analysis of a galaxy cluster gives the
sum of the Bardeen potentials plus negligible corrections.

\subsection{$f(R)$ gravity}

\label{sec:fr} One of the simplest and most investigated alternatives
to General Relativity (GR) is the class of scalar-tensor theories
known as $f(R)$ gravity proposed by ref. \cite{Buch01}, in which
the Einstein-Hilbert action is modified by adding a general non-linear
function of the Ricci curvature scalar $R$. In the Jordan frame the
total action (in units $c=1$) reads: 
\begin{equation}
S\,=\,\frac{1}{16\pi G}\int{\sqrt{-g}[R+f(R)]d^{4}x}+S_{m}[\Psi_{m},g_{\mu\nu}],\label{eq:fr}
\end{equation}
where $S_{m}$ is the action of the matter field $\Psi_{m}$. Note
that for $f(R)=\mathrm{const}\equiv-2\Lambda$ we recover GR in presence
of a cosmological constant.

Variation of eq. \eqref{eq:fr} with respect to the metric $g_{\mu\nu}$
gives rise to the modified Einstein equations: 
\begin{equation}
(1+f_{,R})R_{\mu\nu}-\frac{1}{2}g_{\mu\nu}[f(R)+R]+(g_{\mu\nu}\Box-\nabla_{\mu}\nabla_{\nu})f_{,R}=8\pi GT_{\mu\nu}.\label{field}
\end{equation}
The quantity 
\[
f_{,R}=\frac{df(R)}{dR},
\]
usually known as the scalaron, is a new degree of freedom which can
be interpreted as a scalar field, mediating an additional fifth force
with a characteristic range described by the physical Compton length
$\lambda$ (see below).

In the case of flat FLRW universe, the Ricci scalar is given by 
\begin{equation}
R_{b}=6\left(\frac{\ddot{a}}{a}+H^{2}\right)=3H_{0}^{2}\Omega_{m}\left[(1+z)^{3}+4\frac{\Omega_{\Lambda}}{\Omega_{m}}\right],
\end{equation}
where the subscript $b$ indicates background value, $a(t)$ is the
scale factor, $\Omega_m$ is the matter density parameter today, and $H=\dot{a}(t)/a(t)$ is the Hubble parameter; overdot
means derivative with respect to the cosmic time $t$. Under the assumption
of the cosmological principle, from eq. \eqref{field} we can further
derive the Friedmann equation: 
\begin{equation}
H^{2}+\frac{1}{6}f(R)-\frac{\ddot{a}}{a}f_{,R}+H\dot{f}_{,R}=\frac{8\pi G}{3}\rho_{m,b}.\label{modfr}
\end{equation}
The trace of eq. \eqref{field} shows explicitly the role of the field
$f_{,R}$. Indeed we have: 
\begin{equation}
\Box f_{,R}=\frac{1}{3}\left[R-f_{,R}R+2f(R)-{8\pi G}\rho_{m}\right],\label{eq:motion}
\end{equation}
which could be seen as the equation of motion for the scalaron, with
a canonical kinetic term and an effective potential 
\begin{equation}
\frac{\partial V_{eff}}{\partial f_{,R}}=\frac{1}{3}\left[R-f_{,R}R+2f(R)-{8\pi G}\rho_{m}\right].\label{veff}
\end{equation}
For the class of viable models that in the high-curvature regime satisfy
$|f_{,R}|\ll1$ and $|f(R)/R|\ll1$ (see e.g. refs. \cite{Cataneo16,Pogosian10}),
$V_{eff}$ has an extremum at the general-relativistic value 
\[
R=\frac{8\pi G}{3}\rho_{m}.
\]
The concavity of the potential in the extremum is given by its second
derivative: 
\begin{equation}
m_{f_{R}}^{2}\equiv\frac{\partial^{2}V_{eff}}{\partial f_{,R}^{2}}=\frac{1}{3}\left(\frac{1+f_{,R}}{f_{,RR}}-R\right),\label{mass}
\end{equation}
where $f_{,RR}=df_{,R}/dR$; $m_{f_{R}}$ represents the scalaron
mass and its inverse $\lambda=1/m_{f_{R}}$, the Compton length, gives
the typical interaction range of the fifth force. In the limit of
$|Rf_{,RR}|,|f_{,R}|\ll1$ one has 
\begin{equation}
m_{f_{R}}^{2}\sim\frac{1}{3f_{,RR}}
\end{equation}
Constraining the scalaron mass means therefore to
constrain the second derivative of $f(R)$, rather than just $f_{,R}$.
Notice also that, in this approximation, in order to have a stable
minimum, $f_{,RR}\ge0$ is thus required. Ref. \cite{Song2007} argued
that this is a critical constraint to avoid short timescale instabilities
in the high curvature regime. Physically the condition means a non-tachyonic
scalaron field.

Looking to perturbations in the FRW background, under the quasi-static
approximation for which $\nabla f_{,R}\gg\dot{f}_{,R}$ (which
is achieved for scales $k/aH\gg1$), it is possible to rearrange
the field equation \eqref{eq:motion} in a Poisson-like form for the
fluctuations, as shown in \cite{Oyaizu08}: 
\begin{equation}
\nabla^{2}\delta f_{,R}=\frac{1}{3}\delta R(f_{,R})-\frac{8}{3}\pi G\delta\rho_{m},
\end{equation}
where we are working in physical coordinates and the perturbed quantities
are defined as $\delta X=X-X_{b}$. Solving the linearized modified
Einstein equations in the Newtonian gauge of eq. \eqref{eq:NewGauge}
we can furthermore derive the Poisson equation for the Bardeen potential
$\Phi$: 
\begin{equation}
\nabla^{2}\Phi=\frac{16\pi G}{3}\delta\rho_{m}-\frac{1}{6}\delta R(f_{,R}).\label{eq:pot}
\end{equation}
In the linear regime, the curvature perturbations are everywhere small compared
to the GR value $\delta R\ll8\pi G\delta\rho_{m}$. This happens for example if $|f_{,R(b)}|\gg|\Phi_{N}|$, where
$\Phi_{N}\sim10^{-5}$ is the typical Newtonian potential for a galaxy
cluster. Following e.g. \cite{Cataneo16} and references therein we can expand the fluctuations as: 
\begin{equation}
\delta R\simeq\left(\frac{\partial R}{\partial f_{,R}}\right)_{R_{b}}\delta f_{,R}=3\bar{m}_{f_{R}}^{2}\delta f_{,R},\label{eq:Rlin}
\end{equation}
where $\bar{m}_{f_{R}}^{2}$ refers to the background scalaron mass.
Combining the last three equations and writing the result in Fourier
space, we obtain an expression for the time-time Bardeen potential:
\begin{equation}
\Phi=-\frac{4\pi G\delta\rho_{m}}{1+f_{,R}}\frac{1}{k^{2}}\left(1+\frac{1}{3}\frac{k^{2}}{\bar{m}_{f_{R}}^{2}+k^{2}}\right),\label{phifr}
\end{equation}
which is the same as eq. \eqref{eq:phi} expliciting $Y(k,a)$ with
the assumption: 
\begin{equation}
2Q^{2}=1/3,\,\,\,\,h_{1}=(1+f_{,R})^{-1}\simeq1,\,\,\,\,m={\bar{m}}_{f_{R}}.\label{parfr}
\end{equation}
A similar equation can be derived for $\Psi$: 
\begin{equation}
\Psi=-h_{1}h_{2}\frac{4\pi G\delta\rho_{m}}{k^{2}}\left(1-2\hat{Q}^{2}\frac{k^{2}}{\bar{m}_{f_{R}}^{2}+k^{2}}\right)=-\frac{4\pi G\delta\rho_{m}}{1+f_{,R}}\frac{1}{k^{2}}\left(1-\frac{1}{3}\frac{k^{2}}{\bar{m}_{f_{R}}^{2}+k^{2}}\right),\label{psifrr}
\end{equation}
where $2\hat{Q}^{2}=-1/3\,,h_{2}=1$. In real space, for a NFW mass
density profile, we finally get: 
\begin{align}
 & \Phi(r)=\left[G\int_{r_{0}}^{r}{\frac{dx}{x^{2}}M(x)} +\frac{1}{3}\phi_{mg}(r,\bar{m}_{f_R})\right]+O(f_{,R});\label{potfr}\\
 & \Psi(r)=\left[G\int_{r_{0}}^{r}{\frac{dx}{x^{2}}M(x)} -\frac{1}{3}\phi_{mg}(r,\bar{m}_{f_R})\right]+O(f_{,R}).\label{psifr}
\end{align}
We are now left with only one free parameter, i.e. the background
scalaron mass $\bar{m}_{f_{R}}$ (or, equivalently, the interaction
range $\lambda$), related to the background $f_{,R}$ through eq. \eqref{mass}.
The maximum enhancement of gravity due to the effect of the fifth
force is $1/3$ with respect to GR on scales $k\gg\bar{m}_{f_{R}}$.
Equation \eqref{potfr} is what we are going to use in the analysis of
the cluster dynamics in order to constrain $\lambda=1/m$.

It is straightforward to compute the lensing potential form eqs. \eqref{potfr},
\eqref{psifr}: 
\begin{equation}
\Phi_{lens}(r)=\frac{1}{2}(\Phi+\Psi)=G\int_{r_{0}}^{r}{\frac{dx}{x^{2}}M(x)}+O(f_{,R}).\label{lens}
\end{equation}
In $f(R)$ models photons perceive only the Newtonian part of gravity
except for a correction of order $\sim f_{,R}$. Thus, for models
with $|f_{,R}|\ll1$, geodesics of photons are unchanged by the presence
of the new degree of freedom.{} This feature is physically
related to the property that $f(R)$ and scalar-tensor theories can
be generated by a conformal rescaling of the metric, together with
the conformal invariance of electromagnetism (see e.g. \cite{Schmidt10}).

For small field values $|f_{,R(b)}|\ll|\Phi_{N}|$, the characteristic
mass becomes larger and the contribution of the force modification
is suppressed. Moreover, if we consider an overdensity such as a galaxy
cluster (assumed to be spherically symmetric), in the interior the
field is close the minimum of the effective potential, given by the
GR limit $\delta R\simeq8\pi G\delta\rho_{m}$; in this case field
gradients are negligible except for a shell at the boundary where
the overdensity matches the cosmological background. The thickness
of this shell is given by 
\begin{equation}
\Delta r=r_{match}-S.\label{thick}
\end{equation}
In the above equation, $r_{match}$ indicates the boundary
of the overdensity and $S$ is dubbed the \textit{screening radius}.
For $r<S$ gravity is described by standard GR and eq. \eqref{eq:pot}
becomes the usual Poisson equation for the Newtonian potential.

In order to use the same formalism of eq. \eqref{phifr}, derived
by linearizing the curvature perturbations, inside the
overdensity, we have to replace $\bar{m}_{f_{R}}$ with an effective
mass $m_{eff}(k)$, which depends on the environmental density, and
is related to the scalaron minimum value inside the overdensity. If
in the region enclosed within the screening radius $S$ we have full screening, $m_{eff}$
becomes so large that the field does not propagate and the additional
terms in eqs. \eqref{phifr},\eqref{psifrr} tend to zero. The mechanism
to recover GR in high density regions by using an environment-dependent
field is known as the chameleon mechanism (ref. \cite{K2004}).

Assuming a constant value for $m_{eff}$ in this picture (i.e. neglecting
the dependence on the environment) is equivalent to assume a screening
mechanism so efficient (or so inefficient) that the transition of the scalaron field to
its background value takes place at scales much larger (much smaller) than the cluster
size.

\section{The MAMPOSSt method}
\label{sec:MAM}
\emph{MAMPOSSt (Modelling Anisotropy and Mass Profiles of Observed
Spherical Systems)} is a method to derive mass profiles of galaxy clusters
from the analysis of the dynamics of the member galaxies under the
assumption of spherical symmetry. In this section we describe the code, developed by ref. \cite{Mamon01} and the modification we have made for the investigation of modified gravity models.

The \emph{MAMPOSSt} method performs a Maximum-Likelihood fit to the
distribution of the galaxies, assumed to be collisionless tracers of the gravitational potential, in the projected phase space $(R,v_{z})$, where $R$ is
the projected radius from the cluster center and $v_{z}$ is the velocity
along the line of sight (LOS), assuming a Gaussian 3D velocity distribution\footnote{\emph{MAMPOSSt} can be generally performed given any model of the 3D velocity distribution. The choice of a Gaussian is the most simple,
but the code has been extensively tested to work quite well on halos drawn from cosmological simulations (see ref. \cite{Mamon01})}.\\
The code solves the spherical Jeans' equation: 
\begin{equation}
\frac{d\left[\nu(r)\sigma^2_{r}(r)\right]}{dr}+2\beta(r)\frac{\nu(r)\sigma_{r}^{2}(r)}{r}=-\nu(r)\frac{d\Phi(r)}{dr},\label{eq:Jeans}
\end{equation}
 given a parametric form of the mass profile $M(r)$, the velocity
anisotropy profile $\beta(r)=1-{\sigma_{\theta}^{2}}/{\sigma_{r}^{2}}$ (where $\sigma_\theta$ is the velocity dispersion along the tangential direction),
and the number density profile of the galaxies $\nu(r)$.
In eq. \eqref{eq:Jeans}  $\Phi(r)$ is the time-time Bardeen potential, that coincides with
the Newtonian potential in standard GR, and $\sigma_r(r)$ is the velocity
dispersion along the radial direction.

The solution of the Jeans' equation can be written as (e.g. ref.
\cite{MamLok05}): 
\begin{equation}
\sigma^2_r(r)=\frac{1}{\nu(r)}\int_{0}^{\infty}\exp\left[2\int_{r}^{s}\beta(t)\frac{dt}{t}\right]\nu(s)\frac{d\Phi(s)}{ds}ds.\label{eq:sigmar}
\end{equation}
Note that eq. \eqref{eq:sigmar} is the only relation in which the
gravitational potential enters directly; we thus modified this expression
by substituting $d\Phi/ds$ with the derivative of eq. \eqref{eq:phi}.
The solution is obtained in GR by assuming that the cluster is an
isolated object. Ref. \cite{Falco13} showed that eq. \eqref{eq:sigmar}
holds also in a cosmological $\Lambda$CDM background; since the additional
contribution of fifth force tends to zero when the density reaches
the background value and since we are looking for models that mimic
 the $\Lambda$CDM expansion history, we can safely assume that
the solution of eq. \eqref{eq:Jeans} is still valid in modified gravity.

The probability density of observing an object at position $(R,v_{z})$
in the projected phase space is given by: 
\begin{equation}
q(R,v_{z})=\frac{2\pi Rg(R,v_{z})}{N_{p}(R_{max})-N_{p}(R_{min})},\label{q}
\end{equation}
with $N_{p}(R)$ the predicted number of galaxies with projected radius
$R$ and $g(R,v_{z})$ is the surface density of observed objects with LOS velocity $v_z$ that, in the case of a 3D-Gaussian velocity distribution, takes the
form: 
\begin{equation}
g(R,v_{z})=\sqrt{\frac{2}{\pi}}\int_{R}^{\infty}\frac{r\nu(r)}{\sqrt{r^{2}-R^{2}}}\frac{dr}{\sigma_{r}(r)\sqrt{1-\beta(r)R^{2}/r^{2}}}\exp\left[-\frac{v_{z}^{2}}{2(1-\beta(r)R^{2}/r^{2})\sigma_{r}^{2}(r)}\right].\label{eq:grv}
\end{equation}
The steps of the \emph{MAMPOSSt} procedure could be summarized as
follows:\\
\begin{description}
\item  (i) For a given choice of the parameter vector $\bm{\theta}$, the
code computes $\sigma_{r}(r)$ over a logarithmic grid of points $r_{i}$
and then performs a cubic-spline interpolation to evaluate for intermediate
radii.
 \item (ii) The solution of the Jeans equation is then used to get $g(R,v_{z})$
for each $(R_{i},v_{z,i})$. The integral of eq. \eqref{eq:grv} is
numerically solved by assuming a cutoff for the upper limit of $\sim15r_{200}$,
where the velocity of the matter pushed by the Hubble flow is roughly
$3\sigma$ above the mean value of the cluster. Variations of the
viral radius by up to a factor of 2 do not change significantly the result of the integration
 (ref. \cite{Mamon01}).
\item  (iii) The Likelihood is computed from eq. \eqref{q} by: 
\end{description}
\begin{equation}
-\log\mathcal{L}=-\sum_{i=1}^{n}\log q(R_{i},v_{z,i}|\bm{\theta}).\label{lik}
\end{equation}
The minimum value of $-\log\mathcal{L}$ is found by searching over a grid in the parameter space. The implemented original version of \emph{MAMPOSSt} can work with four free parameters, namely the scale radius of the tracers density profile
$r_{\nu}$, the parameter of the velocity anisotropy profile,
the scale radius $r_{s}$ and the radius $r_{200}$ of the chosen
parametric mass profile. We have expanded the parameter array in the
case of modified gravity analysis by including 3 additional parameters,
which are the interaction range $\lambda=1/m$, $h_{1}$ and the screening
radius $S$. In general we should consider also $Q$ (see Sect. \ref{sec:gen}),
but since we are looking at $f(R)$ models as a case study, in our
modified \emph{MAMPOSSt} code $Q$ is a constant fixed such that $2Q^{2}=1/3$.
Furthermore, $h_{1}=1/(1+f_{,R})$; we restrict our analysis to the
range $\lambda\le100\mpc$, which, for example, in the Hu \& Sawicki
model of $f(R)$ (ref. \cite{Hu2007}) roughly corresponds to $|f_{,R}|<10^{-3}$,
so we can safely set $h_{1}=1$. We will upgrade the method for
running with generic MG theories and other choices of the mass profile
in future works.

\begin{figure}
\includegraphics[width=1\textwidth]{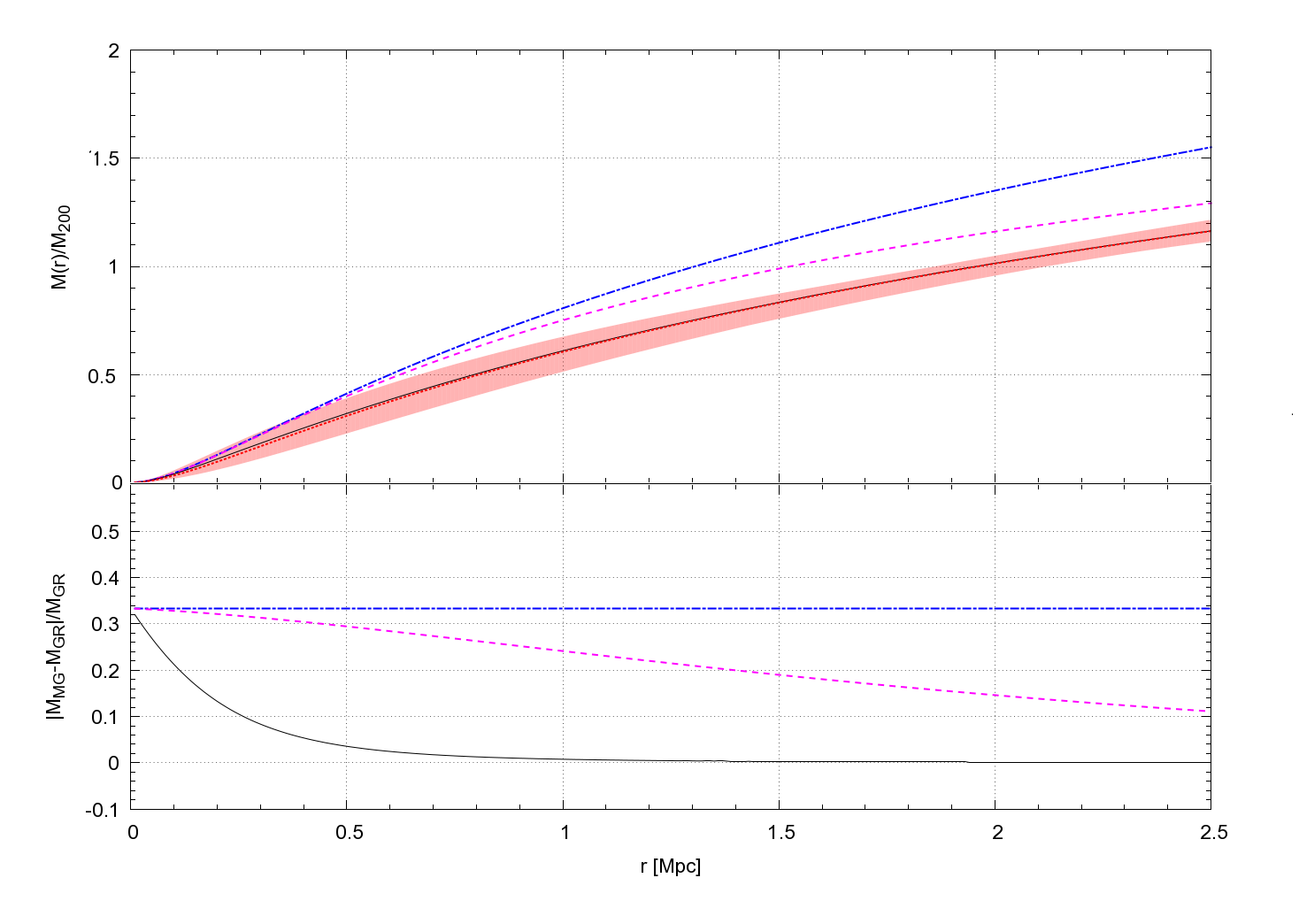} \caption{\label{fig:profile} Upper panel: mass profiles in $f(R)$ gravity,
expressed in unit of $M_{200}=200H^{2}(z)r_{200}^{3}/2G$, for different
values of the interaction range and $r_{200}=1.96\,\mpc$, $r_{s}=0.27\,\mpc$
(best fit values of the GR analysis of ref \cite{Biviano01} for MACS
1206). Blue line-dotted curve: $\lambda=1000\,\mpc$,. Purple dashed
curve: $\lambda=1\,\mpc$. Black solid curve: $\lambda=0.1\,\mpc$.
The red shaded area shows the GR profile within the 68\% C.L. in the
NFW parameter with the best fit indicated by the red dashed curve.
Lower panel: absolute enhancement with respect to the GR best fit
profile. The lines correspond to the same values of $\lambda$ as
in the upper plot. For $\lambda\gg1\,\mpc$ the profile is enhanced by a
factor 1/3 with respect the GR value, while for $\lambda=0.\,\mpc1$ the
result is very close to standard gravity.}
\end{figure}

We tested this modified-\emph{MAMPOSSt} code in the limit of standard
gravity (i.e. $\lambda\to0$) to reproduce the results of ref. \cite{Biviano01}
for MACS 1206; in the opposite situation ($\lambda\gg1$)
we checked that the modification in the mass profile reaches the
maximum enhancement of $1/3$ as expected for $f(R)$ gravity (see
Fig \ref{fig:profile}).\\

\section{MACS 1206 and RXJ 2248}
\label{sec:cluster}
The galaxy clusters MACS 1206, at redshift $z=0.44$, and RXJ 2248, at redshift $z=0.35$, belong to a sample of 20 X-ray selected clusters for their apparent properties of dynamical relaxation, analysed within the CLASH project (ref. \cite{Postman01}) and its spectroscopic follow-up CLASH-VLT (ref. \cite{Rosati1}). 

The analyses of ref. \cite{girardi15}  and ref. \cite{Lemze01} confirmed that MACS 1206 is a relaxed system with minor overdensities in the two-dimensional distribution and a negligible level of substructures within the cluster when the most conservative membership selection is used.  Moreover, the mass profile derived by the \emph{Chandra} X-ray data under the assumption of hydrostatic equilibrium of the intra-cluster medium is in agreement with the mass profile obtained by the Jeans' analysis of the dynamics of the galaxies inside the cluster (ref. \cite{Biviano01}). Since the two determinations are both sensitive to the time-time Bardeen potential $\Phi$,  but the equilibrium is reached in different ways, the consistency between the two results is another indication of the relaxed state of the cluster. The dynamical relaxation is further suggested by the concentric distribution of the stellar, gas and dark matter mass components, as argued in ref. \cite{UmetsuMACS}.

The kinematic determination of the mass profile was carried on by ref. \cite{Biviano01} which used spectroscopy information from the CLASH-VLT
project (ref. \cite{Rosati1}). From the sample of 2749 galaxies observed with VLT/VIMOS with reliable redshift measurements in the cluster field, 592 cluster members were identified after the rejection of interlopers. 

Ref. \cite{Biviano01} applied the \emph{MAMPOSSt} method in the phase-space of the member galaxies to estimate the mass profile up to the virial radius where the Jeans' equation is supposed to be valid. Three models for the anisotropy profile $\beta(r)$ have been taken into account, namely the Tiret model
\begin{equation}
\beta_{T}(r)=\beta_{\infty}\frac{r}{r+r_{c}},\label{eq:tmod}
\end{equation}
from ref. \cite{Tiret01}, hereafter denoted as the "T" model, the modified Tiret model
\begin{equation}
\beta_{O}(r)=\beta_{\infty}\frac{r-r_{c}}{r+r_{c}},\label{eq:omod}
\end{equation}
denoted as "O" model, and a model with constant anisotropy $\beta(r)=\beta_C$ (the "C" model). In eqns. \eqref{eq:tmod}, \eqref{eq:omod} $\beta_{\infty}$ represents the
anisotropy value at large radii which is the free parameter entering in the \emph{MAMPOSSt} analysis. $r_c$ is assumed to coincide with the scale radius $r_s$ of the mass profile. This choice, as shown in ref. \cite{Mamon10}, provides a good fit to the average anisotropy profiles of  galaxy clusters in a set of cosmological simulations.
Similarly, ref. \cite{Biviano01} considered different parametric expressions for the mass profile: the Einasto model (ref. \cite{Einasto65}), the NFW model,  the Hernquist model (ref. \cite{Hernquist01}), the Burkert model (ref. \cite{Burkert01}) and the softened isothermal sphere model (see e.g. ref. \cite{Geller99}).
In the \emph{MAMPOSSt} procedure the maximum value of the likelihood is obtained for the NFW profile; the combination of this mass model and the modified Tiret \textquotedbl{}O\textquotedbl{} anisotropy model gives the smallest product of the relative errors in $r_s$ and $r_{200}$ once marginalized over $\beta$.

Ref. \cite{Zitirn01} performed a first strong lensing analysis for MACS 1206 using 50 multiple images of 13 background sources, further upgraded by ref. \cite{UmetsuMACS}, which combined strong lensing measurements with weak lensing shear and magnification information from \emph{Subaru}
multi-band images out to $\sim 2 \,\mpc$.
The resulting mass profile is parametrized as a NFW profile Additional lensing analyses of the CLASH clusters involving also MACS 1206 were carried out by refs. \cite{2015M,Umetsu14,Zitrin2015,Umetsu16}; in this paper we refer to the results of ref. \cite{Umetsu16}, that refined the joint shear-and-magnification weak-lensing analysis of ref. \cite{Umetsu14} by including {\em HST} strong-lensing information for a sample of 20 CLASH clusters. The weak+strong lensing results take into account model-dependent
systematics of their strong-lensing modelling. Their error analysis
also accounts for the intrinsic variations of the projected cluster
lensing signal due to variations in cluster asphericity and the
presence of correlated halos (see ref. \cite{Zitrin2015}).

In the case of RXJ 2248  (first identified as Abell S1063 in ref. \cite{Abell89}), given its high mass and relatively high redshift, many lensing analyses have been performed:
for the strong lensing analysis refs. \cite{Monna14,Johnson14,Richard14,Zitrin2015,Caminha16}, and for the weak lensing
analyses  refs. \cite{Gruen13,Umetsu14,Merten15,Melchior15,Umetsu16}.
The strong lensing mass profile used in this paper is an improvement
of the one presented in \cite{Caminha16}. They used a data set of 47 multiple images belonging to 16 families in
a redshift range of $[0.1-6]$ detected down to $m_{F814W}=26$. This data-set comes
from the CLASH survey HST imaging, the CLASH-VLT spectroscopy with VIMOS-VLT,
and obtained during the MUSE-VLT from MUSE science verification programme  (ID
60.A-9345, P.I.:K. Caputi).
As we did in the case of MACS1206, we use the weak lensing analysis of ref. \cite{Umetsu16} that combine weak lensing shear and magnification information from WFI
2.2m images out to $\sim2.5\,\mpc$. We combine the weak lensing information with the
strong lensing ones from ref. \cite{Caminha16}. The combined mass profile is
parametrized with a NFW model.

As for the kinematic analysis we used the results obtained in Sartoris et al (in
prep.) In this analysis they use a sample of 1233 galaxy members selected among
more than 3700 galaxies with spectroscopic redshift provided by the VIMOS and
MUSE observation out to $\sim5\,\mpc$. The MAMPOSSt technique has been applied using
several anisotropy and mass profile models. Sartoris et al. in prep find that
the best combination of models is the NFW mass profile with a Tiret velocity
anisotropy model.

According to ref. \cite{Gomez12}, RXJ 2248 has undergone a recent off-axis
merger. However, moderately deep X-ray Chandra (ref. \cite{Caminha16})
observations do not show evidence of massive substructures in the inner region,
but only a regular elongated shape, orientated like a large scale filament
(ref. \cite{Melchior15}).

\section{Results}

\label{sec:results} In this section we present our results for the
constraints on the interaction range $\lambda$ obtained from the
analysis of the galaxy clusters MACS 1206 and RXJ 2248.

We apply the \emph{MAMPOSSt} method to constrain $\lambda=1/m_{eff}$ using the parametric expression of eq. \eqref{eq:phimod} for the gravitational
potential in generic $f(R)$ models (i.e. setting the parameters as
in eq. \eqref{parfr}) without screening, i.e. by assuming that the
dependence of the environment of $m_{eff}$ is negligible at the scales
we are looking. In other words, we assume the screening radius to be much smaller - or much larger - than the cluster size. The situation could be also described
in terms of a model with a strong screening mechanism where the effective
mass of the scalaron remains always close to the minimum inside the
overdensity.

For both clusters we include in the analysis data out to the virial radius, that is
close to $r_{200}$, to ensure the validity of the Jeans equation.
Moreover, the region below $r=0.05\,Mpc$ is excluded since the internal
dynamics of the Brightest Central Galaxy (BCG) becomes dominant (see
e.g. ref. \cite{Biviano06}).

\subsection{MACS 1206}

\label{sec:noscreenM} In the case of MACS 1206, for which 592 cluster
members were identified in ref. \cite{Biviano01}, we use a sample
of 345 galaxies, namely all the members lying within the radial range used in our analysis $[0.05\,\mathrm{Mpc}-2.0\,\mathrm{Mpc}]$ in which we assume the Jeans' equation to be valid. The projected number density
profile of the tracers $n(R)$ is fitted with a projected NFW (pNFW,
ref. \cite{Bartelmann96}) with a face value of the scale radius parameter $r_{\nu}=0.63\,\mpc^{+0.11}_{-0.09}$,
as given by the Maximum Likelihood fit on the total sample of spectroscopic+photometric
members (see Sect. 2.2 of ref. \cite{Biviano01}). 

We run the \emph{MAMPOSSt}
procedure in the modified gravity scenario with four free parameters, $r_{200},\,r_s,\,\lambda$ plus the velocity anisotropy parameter $\beta$. 
We use the \textquotedbl{}C\textquotedbl{} velocity anisotropy model (constant anisotropy with radius) as the reference model of our analysis, since it provides the highest probability in the \emph{MAMPOSSt} fit
in agreement with the GR results (see Table 2 of ref. \cite{Biviano01})


\begin{figure}
\includegraphics[width=1\textwidth]{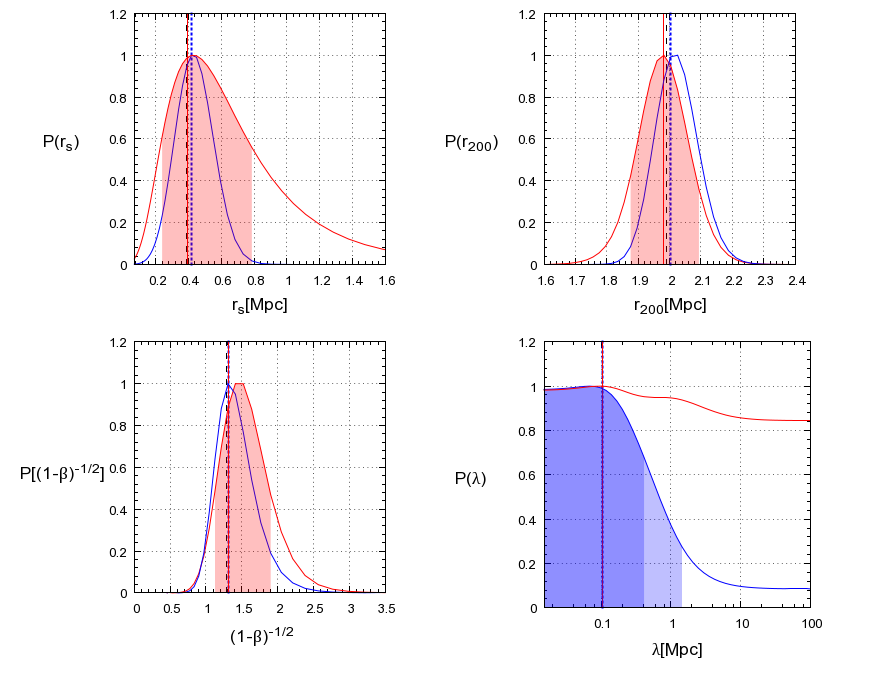} \caption{\label{fig:Macs} Results for MACS 1206. Likelihood distributions for the free parameters
in the \emph{MAMPOSSt} analysis obtained by marginalizing over the
other three. Upper panels: $r_{s}$ and $r_{200}$. Bottom
panels:  $\sigma_r/\sigma_{\theta}=1/\sqrt{1-\beta}$ and $\lambda=1/m_{eff}$. The red distributions are obtained from the kinematic analysis alone, while the blue curves show the joint lensing+kinematic results. Red and blue vertical lines represent the values
corresponding to the maximum of the \emph{MAMPOSSt} Likelihood and of $\mathcal{L}_{tot}$ respectively.
Black solid lines are the best fit values of the GR analysis. Filled
Red shaded areas in $r_{s}$, $r_{200}$ and $\beta$ indicate the $68\%$ C.L. error from the dynamic analysis; the dark and light blue regions below the distribution of $\lambda$ show the $\Delta \chi^2=1.0$ and the $\Delta \chi^2=2.1$ confidence intervals respectively.}
\end{figure}

In Figure \ref{fig:Macs} we show the marginalized likelihood distributions
for the four free parameters in the case of $\beta(r)=const$. As in ref. \cite{Biviano01}, we consider the quantity $\sigma_r/\sigma_{\theta}=1/\sqrt{1-\beta}$ instead of $\beta$. 
The red curves are the results from the kinematic analysis while the blue curves represent the marginalized distributions when adding lensing information (see below). The vertical red solid lines indicate the best fit values given by \emph{MAMPOSSt}
for each parameter, compared with the GR best fit (black dashed lines). The red shaded regions in the $r_{s}$, $r_{200}$ and $\beta$
probability distributions show the $68\%$ statistical errors. 
We obtain: 
\begin{equation}
r_{200}=1.98\pm0.11\,\mpc,\,\,\,\,\,\,\,\,\,\,\,\,\,r_{s}=0.39_{-0.16}^{+0.40}\,\mpc,\,\,\,\,\,\,\,\,\,\,\,\,\beta\equiv\beta_C=1.31^{+0.59}_{-0.19},
\end{equation}
at $68\%$ C.L., consistent with the results of \cite{Biviano01}. This implies that the constraints on the standard parameters in GR   are basically unaffected by the introduction of the modified gravity
term in the dynamical analysis with the \emph{MAMPOSSt} method.

From the red curve in the bottom right panel of Fig. \ref{fig:Macs} it can be seen that the
distribution of $\lambda$ from the dynamics alone is almost flat; in the limit of $\lambda\to0$ (i.e. $m_{eff}\to\infty$)
this is not surprising, since we are approaching the GR regime and
the contribution of $M_{mg}$ in eq. \eqref{eq:phimod} becomes negligible.
To give an example, for $\lambda\le0.05\,\mpc$ the increase in the mass
profile at $r=0.3\,\mathrm{Mpc}$ due to the fifth force is $\sim10^{-4}$,
that is undetectable if compared to the modification induced by the
statistical uncertainties in $r_{s}$ and $r_{200}$.

\begin{figure}
\includegraphics[width=1\textwidth]{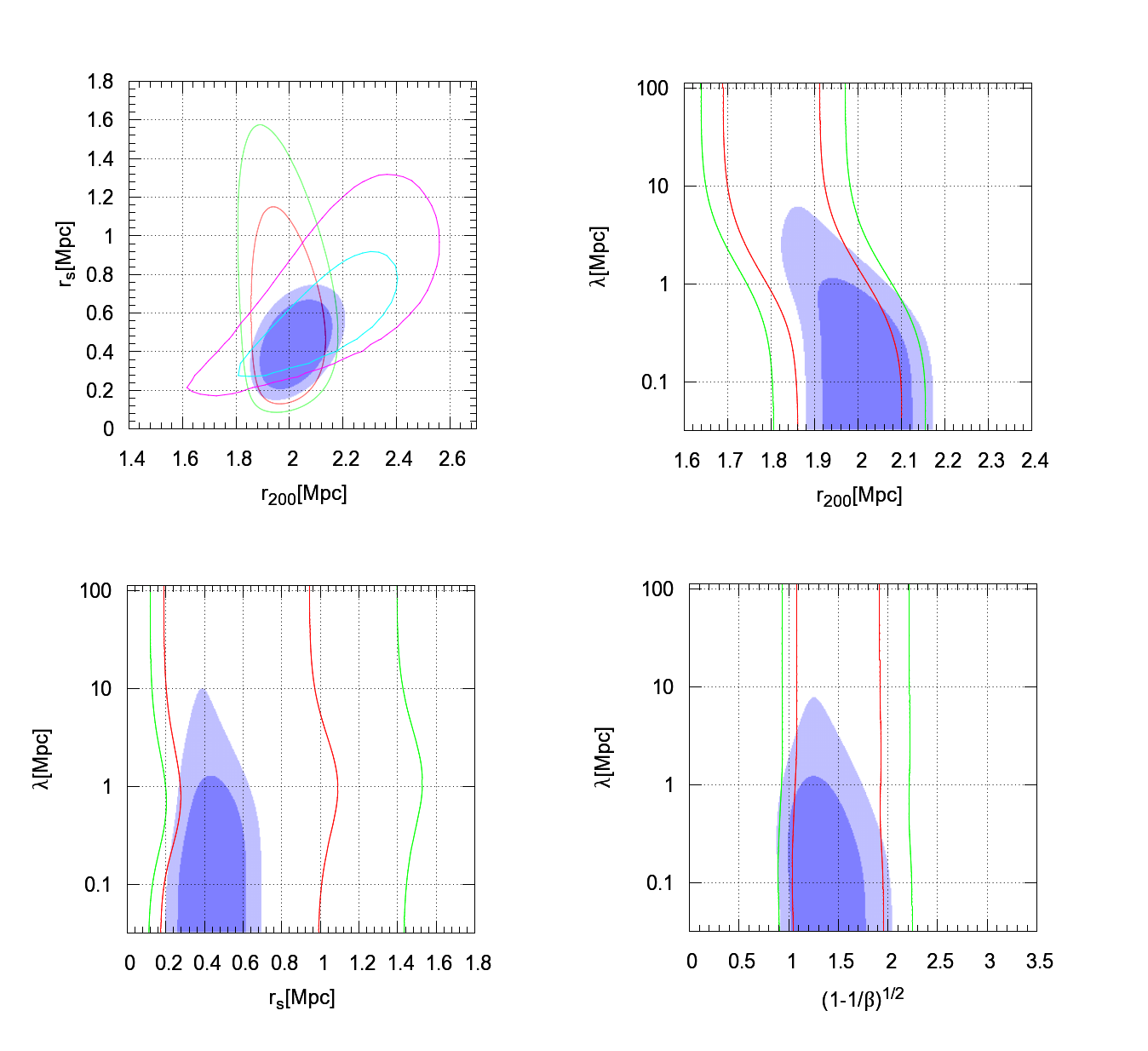} \caption{\label{fig:Macs2} Results for MACS 1206. Two-dimensional likelihood distributions obtained
from the dynamical analysis alone (red and green contours) and from the total likelihood with lensing contribution (dark and light blue shaded areas) after marginalization over the other two parameters.
Upper panels: $r_{200}$ vs $r_{s}$, $r_{200}$ vs $\lambda$.
Bottom panels $r_{s}$ vs $\lambda$, $\beta$ vs $\lambda$. The inner contours/shaded regions correspond
to the points within $\Delta\chi^{2}\le2.3$ from the maximum of
the probability (roughly measuring the 68\% C.L) while the outer contours/filled
regions indicates points within $\Delta\chi^{2}\le4.6$ (which represents
the 90\% C.L.). In the upper left panel the cyan and magenta lines show the $\Delta\chi^{2}\le2.3$ and  $\Delta\chi^{2}\le4.61$
contours from the lensing analysis of ref. \cite{Umetsu16}.}
\end{figure}

More interesting is the case $\lambda\gg1\,\mpc$, associated with considerable
deviations from GR. The flattening behaviour of the curve for large
$\lambda$ is explained by looking at the degeneration directions
in the two-dimensional distributions of Fig. \ref{fig:Macs2}. Each
plot is obtained by marginalizing over the other two parameters; here the red and green lines
 indicate the dynamics contours at $\Delta\chi^{2}=2.3,\,\,4.61$ 
(where $\chi^{2}=-2\log[\mathcal{L}])$ respectively. As $\lambda$ increases,
the NFW parameters tend to assume lower values; this is particularly
evident for $r_{200}$ while $r_{s}$ shows only a slight change for
$\lambda>1$. This behaviour is a consequence of the relatively small maximum gain
(up to $1/3$) produced in the mass profile by the additional force
in $f(R)$ gravity. In fact, the effect generated by the term $M_{mg}$
for large values of $\lambda$ can be compensated  by suitably adjusting
$r_{s}$ and $r_{200}$ with respect to the GR values, so that the
resulting modified mass profile becomes very close to the GR one. Physically, this means that in $f(R)$ gravity the galaxy dynamics is altered in a similar way as the modification induced by a deeper potential well in GR, and even the high-quality data we are using
are not sufficient to distinguish between the two cases. 

The interaction range distribution shows a smooth peak at $\lambda=0.1\,\mpc$, corresponding to the best
fit given by \emph{MAMPOSSt}, but the excess
of probability is statistically irrelevant.

As discussed in Sect. \ref{sec:fr}, up to a conformal rescaling $(1+f_{,R})^{-1}$,
photons are affected only by the Newtonian contribution in $f(R)$
models. We thus can get additional information on the NFW parameters
$r_{s}$ and $r_{200}$ by using the results of the gravitational lensing analysis presented in ref. \cite{Umetsu16}.

In order to improve the derived constraints, we multiply the total likelihood
distribution obtained by the \emph{MAMPOSSt} method with the posterior
probability distribution $P_{lens}(r_{s},r_{200})$ given by strong+weak
lensing analysis of \cite{Umetsu16} assuming flat prior in $\log(M_{200}),\,\,\log(c)$.
Since the lensing distribution peaks at $r_{s}=0.53\pm0.18\,\mpc,\,\,r_{200}=2.14\pm0.16\,\mpc$\footnote{here the errors are estimated by approximating the distribution to a bivariate gaussian around the maximum of the probability. Note that the maximum-likelihood values of  $r_{200}$ and $r_s$ we
find are slightly smaller than the values given in Table 2 of ref.
\cite{Umetsu16}. This is because they quoted marginalized posterior
constraints on the respective parameters obtained using the biweight
location and scale estimators of ref. \cite{Beers90}},
favouring larger values of the NFW parameters compared to the \emph{MAMPOSSt}
result (see left panel of Fig. \ref{fig:Macs2}), the final likelihood,
defined as 
\begin{equation}
\log(\mathcal{L}_{tot})=\log(\mathcal{L}_{dyn})+\log(P_{lens}),\label{eq:l}
\end{equation}
explores a region in the parameter space that is in the orthogonal direction
with respect to the degeneracy direction in the dynamical analysis,
thus increasing the significance of small deviations from GR. 
The new two-dimensional contours at $\Delta\chi^{2}=2.3,\,\,4.61$, obtained by including the lensing prior, are shown by the dark and light blue regions in  Fig. \ref{fig:Macs2}, while the marginalized distributions of $\mathcal{L}_{tot}$ are plotted as the solid blue curves in Fig. \ref{fig:Macs}. The blue vertical dotted lines in each panel correspond to
the maximum of the total likelihood $\mathcal{L}_{tot}$ including lensing informations. Now we can put an upper limit, after marginalization,
on the effective scalaron interaction range $\lambda\le0.49\,\mpc$
at $\Delta\chi^{2}=1.0$ (blue shaded area in the right bottom panel
of Fig. \ref{fig:Macs}) and $\lambda\le1.01\,Mpc$ at $\Delta\chi^{2}=2.71$
(light blue shaded area in the right bottom panel of Fig. \ref{fig:Macs}),
in agreement with the results of ref. \cite{Pizzuti16} which indicate
negligible deviations from GR for this cluster.

\begin{figure}
\includegraphics[width=1\textwidth]{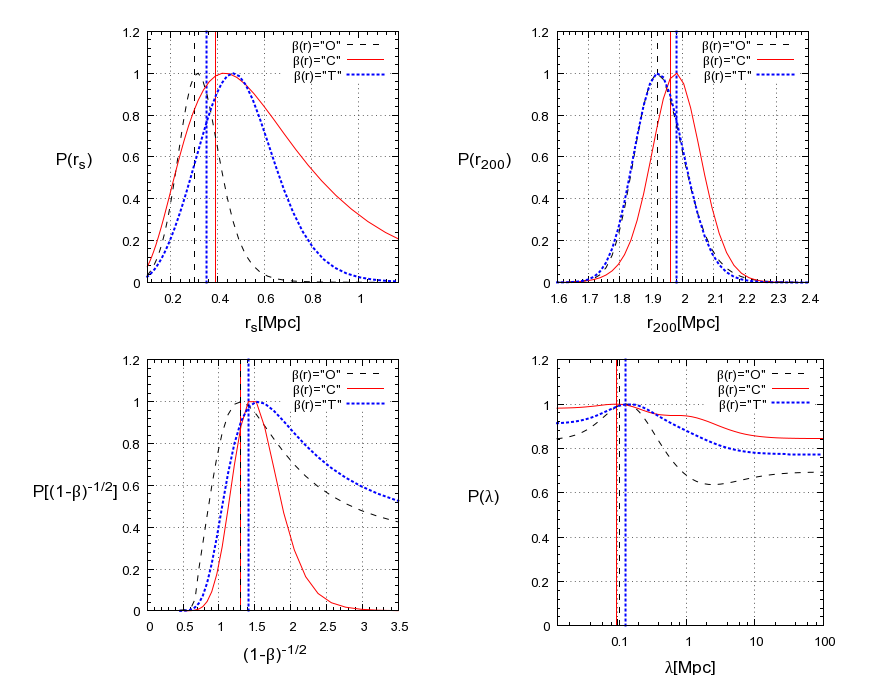}
\caption{\label{fig:sysM_beta} Results for MACS 1206. Marginalized likelihood distributions of $r_{200}$,
$r_{s}$, $\sigma_r/\sigma_{\theta}=1/\sqrt{1-\beta}$ and $\lambda$ from the \emph{MAMPOSSt} analysis obtained
by changing the anisotropy model $\beta(r)$ for $r_{\nu}=0.63\,\mpc$. in the bottom left plot $\beta$ indicates $\beta_C$ for the \textquotedbl{}C\textquotedbl{} model and $\beta_{\infty}$ for \textquotedbl{}T\textquotedbl{} and \textquotedbl{}O\textquotedbl{} models. Black dashed curves: \textquotedbl{}O\textquotedbl{} model. Red solid curves: \textquotedbl{}C\textquotedbl{} model.  Blue dotted curves
:\textquotedbl{}T\textquotedbl{} model. The corresponding vertical
lines indicate the \emph{MAMPOSSt} best fit of each free parameter.}
\end{figure}

\begin{figure}
 \includegraphics[width=1\textwidth]{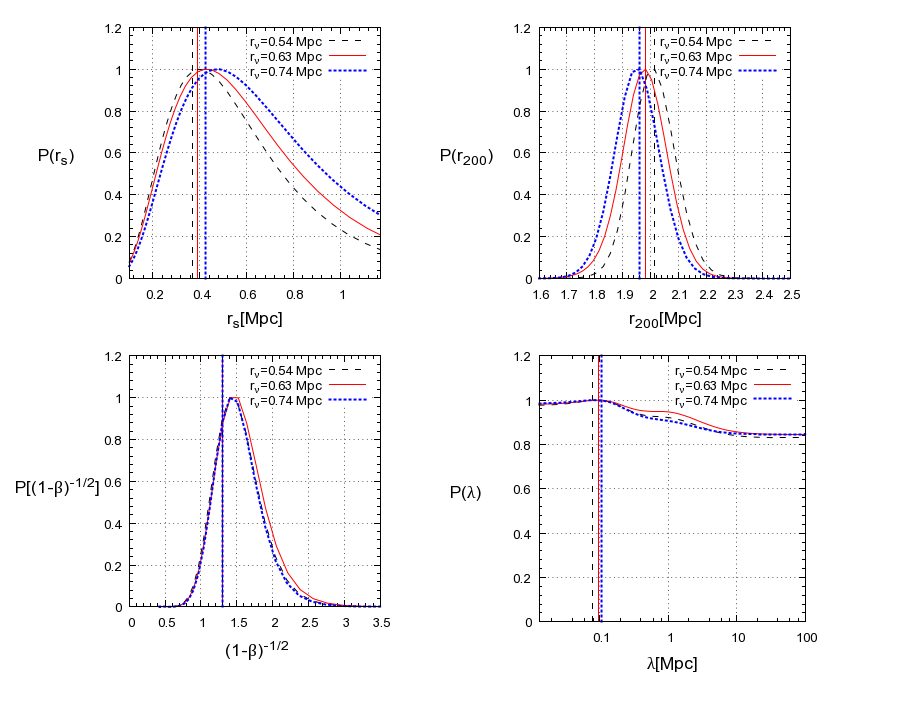}
\caption{\label{fig:sysM_rnu} Results for MACS 1206. Marginalized likelihood distributions of $r_{200}$,
$r_{s}$, $\sigma_r/\sigma_{\theta}=1/\sqrt{1-\beta}$ and $\lambda$ obtained for the reference model \textquotedbl{}C\textquotedbl{}
by changing the scale radius of the number density profile of the galaxies $r_{\nu}$ within the 68\% confidence region of the G.R. analysis. Red solid curves: $r_{\nu}$  fixed to the GR best fit value. Black dashed curves: $r_{\nu}$ fixed to
the lower limit of the 68\% confidence region. Blue dotted curves: $r_{\nu}$ ($\beta_{\infty}$) settled to
the upper limit of the 68\% confidence region. The corresponding vertical
lines indicate the \emph{MAMPOSSt} best fit of each free parameter.}
\end{figure}

The analysis discussed above has been performed for our best fit model, NFW+\textquotedbl{}C\textquotedbl{}, by assuming a fixed value for the scale radius of the number density profile of the tracers $r_{\nu}$
as the best fit value given by ref. \cite{Biviano01}. We can asses
now by how much the constraints are affected by a change of the anisotropy profile and of the parameter $r_{\nu}$ .
To this purpose, we repeat the analysis for the other two anisotropy models mentioned, \textquotedbl{}T\textquotedbl{} and \textquotedbl{}O\textquotedbl{}; moreover, for our reference model we run \emph{MAMPOSSt} with $r_{\nu}$
set to the extremes of the 68\% confidence region obtained by
the GR analysis of ref. \cite{Biviano01}. The results are summarized in the first 5 rows of Table \ref{Tab1} while the marginalized distributions are shown in Fig. \ref{fig:sysM_beta} and Fig. \ref{fig:sysM_rnu}. 

In each plot of both figures the red lines are for
the distributions obtained from the reference model \textquotedbl{}C\textquotedbl{} with $r_{\nu}=0.63\,\mpc$.
In Fig. \ref{fig:sysM_beta}, the black dashed line shows the results for the \textquotedbl{}O\textquotedbl{} model, while the blue dotted curve is obtained from the \textquotedbl{}T\textquotedbl{} model. It is worth to notice that the best fit value of $\lambda$, as well as the position of the peak of the interaction range marginalized distribution, is not modified significantly by the different parametrization of the anisotropy profile. This is not surprising since $\lambda$ is totally degenerate with $\beta$, as we can see from the right bottom plot of Fig. \ref{fig:Macs2}\footnote{We verified that this statement remains valid for the other two models analysed}. On the other hand, the shape of the distribution near the peak is affected by the changes in the anisotropy model. This is a consequence of the degeneration between $\lambda$ and the scale radius $r_s$, which is responsible of the internal structure of the mass profile (right bottom panel of Fig. \ref{fig:Macs2}). In fact, the more the distribution in $r_s$ is sharp, the more the peak in $\lambda$ is evident.
Fig. \ref{fig:sysM_rnu} shows the effect of changing $r_{\nu}$ within the 68\% C.L. given by the GR analysis. The relatively small statistical uncertainties with which $r_{\nu}$ is known produce a negligible effect on the marginalized distributions of the free parameters in our analysis. The black dashed curves in each plot indicate the results for $r_{\nu}$ set at the upper limit of the $68\%$ confidence region, while the blue dotted curves are for $r_{\nu}$ set to the lower value. It's interesting that this last case ($r_{\nu}=0.54\,\mpc$) produce a likelihood slightly higher than the best fit $r_{\nu}=0.63\,\mpc$ (see Table \ref{Tab1}), but we stress again that the scale radius is obtained by a fit 
which is external to the \emph{MAMPOSSt} procedure. The corresponding vertical lines indicate the values which maximize the likelihood $\mathcal{L}_{dyn}$.

To estimate the systematic uncertainties in our results on $r_s$, $r_{200}$ and $\lambda$ we 
take into account the variation caused by these changes. For the standard NFW parameters we obtain: 
\begin{equation}
r_{200}=\left[1.98\pm0.11(stat)\pm0.06(syst)\right]\mpc,\,\,\,\,\,\,\,\,r_{s}=\left[0.39_{-0.16}^{+0.40}(stat)\pm0.09(syst)\right]\,\mpc,
\end{equation}
where the statistical errors indicates the $68\%$ confidence region
as above, and the systematic errors are computed as the maximum difference
between the best fit values (vertical lines). For $\beta\equiv\beta_C$ we can estimate only the variation induced by $r_{\nu}$, since the parameter has a different meaning for each velocity anisotropy profile. As we can see from Fig. \ref{fig:sysM_rnu}, no measurable effects are produced on the $\beta$ distribution when $r_{\nu}$ is modified.

In the case of $\lambda$, we estimate the impact of systematics looking
at the modifications in the upper bound after adding information from
lensing analysis: 
\begin{equation}
\lambda\le[1.61+0.30(syst)]\mpc\,\,\,\,\,\,\,\,\,\,\,\Delta\chi^{2}=2.71,
\end{equation}
where the systematic uncertainty is the largest difference (in absolute value) between the upper limits obtained from the models analysed. The result is still compatible with small or negligible deviations from
standard gravity.

\subsection{RXJ 2248}

\label{sec:noscreenR} We perform our analysis for the cluster RXJ
2248 in the radial range $[0.05\,\mpc-2.3\,\mpc]$, using the 981
member galaxies lying in this region out of a sample of 1233 cluster
members identified (Sartoris et al., in prep.).

\begin{figure}
\includegraphics[width=1\textwidth]{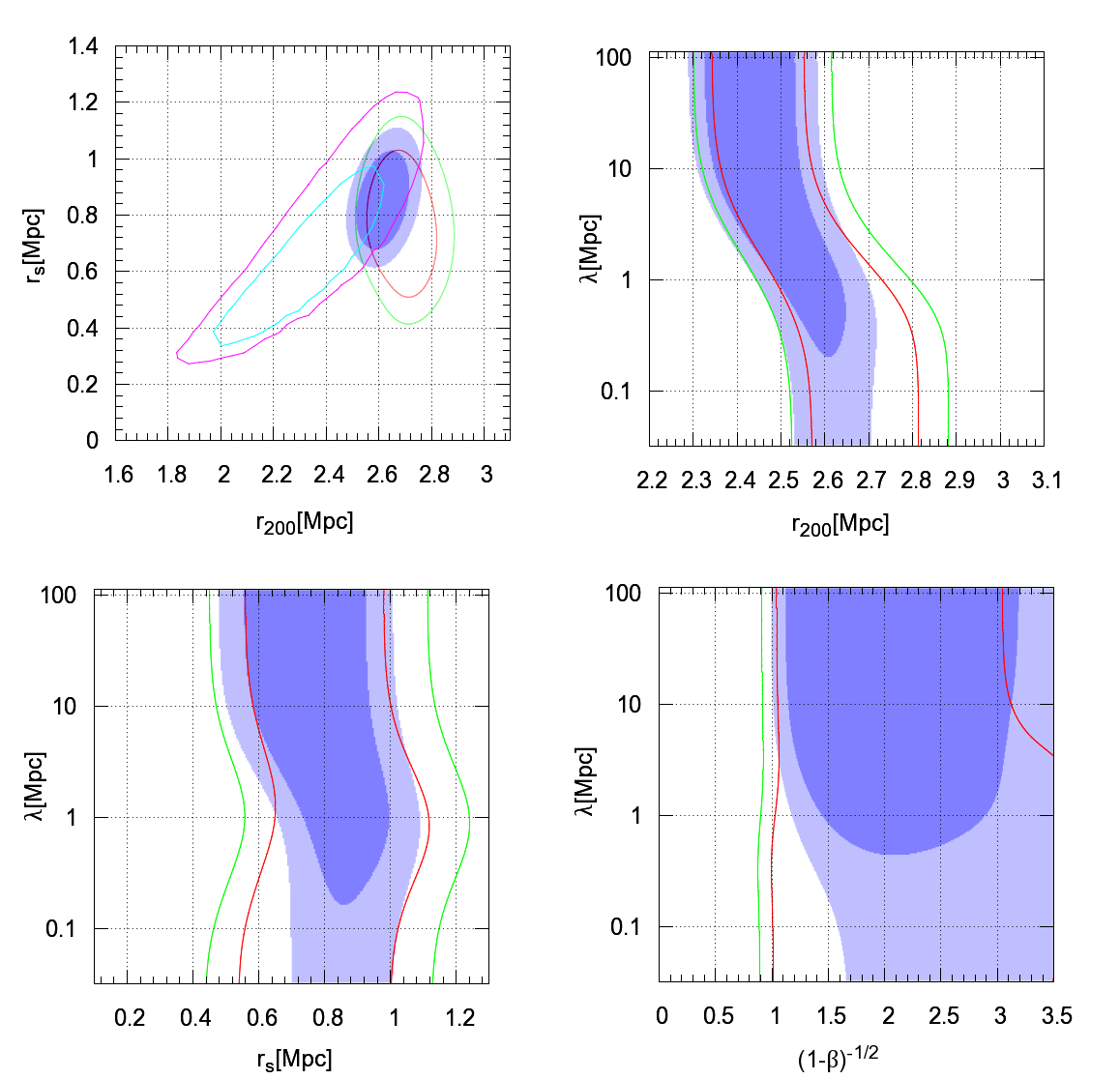} 
\caption{\label{fig:margR2} Results for RXJ 2248:
two-dimensional contours at $\Delta\chi^{2}\le2.3$ and $\Delta\chi^{2}\le4.61$
from the maximum of the probability of the \emph{MAMPOSSt} analysis
(red and green lines respectively) and of the joint dynamics+lensing
analysis (dark blue and light blue filled regions). The cyan and purple  lines
in the left upper panel show the lensing contours at 68\% C.L. and 90\% C.L.}
\end{figure}

\begin{figure}
\includegraphics[width=1\textwidth]{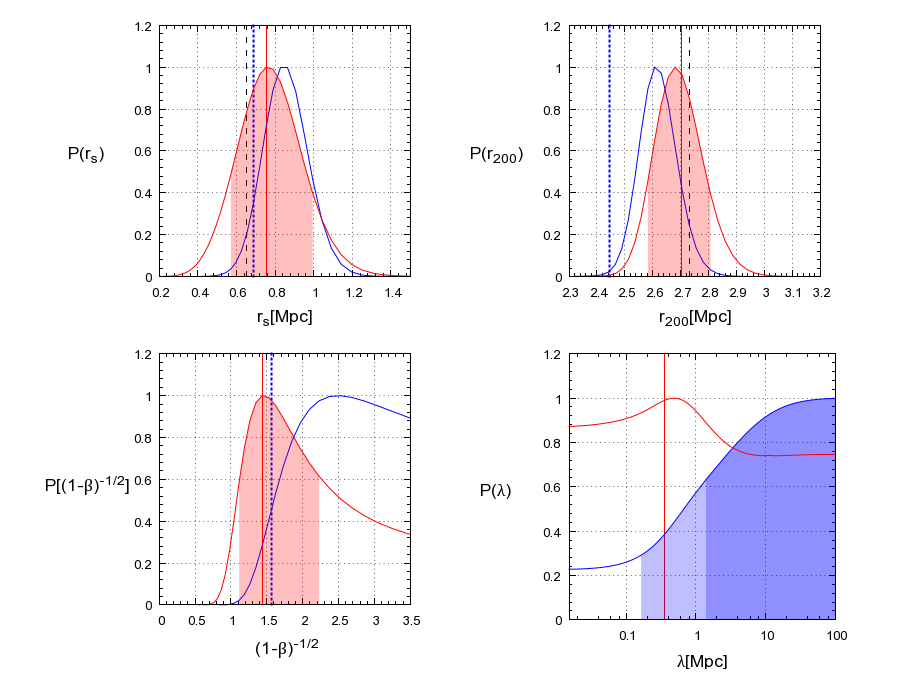} 
\caption{\label{fig:margR}
Results for RXJ 2248: single parameter distributions obtained by marginalizing the
total likelihood over the other three. Upper panels: $r_{s}$ and $r_{200}$. Bottom
panels:  $\sigma_r/\sigma_{\theta}=1/\sqrt{1-\beta}$ and $\lambda=1/m_{eff}$. Red curves: dynamics analysis,
blue curves: lens+dyn. Vertical red solid lines indicate the \emph{MAMPOSSt}
best fit values, blue dotted lines correspond to the values maximizing
the total likelihood while black dashed lines are the dynamics GR
best fit. The filled areas below the marginalized distributions of
$r_{s}$, $r_{200}$ and $\beta=\beta_{\infty}$ represent the 68\% C.L. of the dynamics results.
The dark and light blue regions below the curve $P(\lambda)$ show
the $\Delta\chi^{2}\le1.0$ and $\Delta\chi^{2}\le2.71$ confidence
intervals, respectively.}
\end{figure}

As for MACS 1206, we fixed the number density profile whose parameter is given by the
best fit value of the standard GR study (Sartoris et al., in prep.).
In particular, the projected number density profile is again a pNFW
with $r_{\nu}=0.59\pm0.08\,\mpc$, where this value is obtained by a Maximum Likelihood analysis on the spectroscopic members only. For $\beta(r)$ we use the same models discussed above; the highest probability for this cluster is given by the Tiret model \textquotedbl{}T\textquotedbl{} of eq. \eqref{eq:tmod}.
The results of the \emph{MAMPOSSt} procedure applied to the modified
gravitational potential of eq. \eqref{eq:phimod} with the velocity anisotropy model \textquotedbl{}T\textquotedbl{} are shown in Figs. \ref{fig:margR2} and \ref{fig:margR}. As before,  the red and green curves in Fig. \ref{fig:margR2}  indicate the contours at $\Delta\chi^2=2.3$ and $\Delta\chi^2=4.6$ respectively. The distributions for each parameter, obtained after marginalizing $\mathcal{L}_{dyn}$, are plotted as the red curves in Fig. \ref{fig:margR}.

The effective interaction range probability distribution (bottom right plot of Fig. \ref{fig:margR}) shows the same qualitative behaviour as found from the
analysis of MACS 1206: the curve flattens both for large $\lambda$ and for $\lambda\to0$, although the shape is more peaked near the \emph{MAMPOSSt}
best fit value $\lambda=0.91\,\mpc$ (red vertical solid line) compared to the previous cluster. As mentioned above, the presence of the peak is strictly related to the distribution of the scale radius $r_s$, which in this case is sharper than the one found from the analysis of the reference model for MACS 1206.
Also the 2-dimensional degeneration directions with $r_{s}$, $r_{200}$ and $\beta$ (Fig. \ref{fig:margR2})
are very similar to those of Fig. \ref{fig:Macs2}. The reason for
a nearly constant, non-zero probability to have large deviations
form GR can be again explained in terms of the small enhancement
of the mass profile produced by the fifth force, which allows $\lambda$
to be large if $r_{s}$ and $r_{200}$ change in such a way to compensate
the effect of the additional term. It is worth to point out that the
number of galaxies used in the \emph{MAMPOSSt} fit for this cluster
is about three times larger than that of MACS 1206; this means that roughly tripling
the statistics is still not sufficient to set meaningful constraints on $\lambda$
from the analysis of dynamics alone.

As for the standard NFW parameters and the velocity anisotropy, we obtain the results:
\begin{equation}
r_{200}=2.70_{-0.14}^{+0.10}\,\mpc,\,\,\,\,\,\,\,\,\,\,\,\,\,r_{s}=0.75_{-0.18}^{+0.14}\,\mpc,\,\,\,\,\,\,\,\,\,\,\,\,\beta_{\infty}=1.43^{+0.59}_{-0.19},
\end{equation}
at $68\%$ C.L., where the errors are computed with respect to
the best fit values of the \emph{MAMPOSSt} procedure (red vertical solid lines in Fig. \ref{fig:margR}).
The constraints are consistent with the results obtained from
the GR analysis (Sartoris et al., in prep.) $r_{200}=2.73_{-0.09}^{+0.08}\,\mpc$, $r_{s}=0.65_{-0.22}^{+0.12}\,\mpc$
and $\beta_{\infty}=1.43^{+1.07}_{-0.24}$ (black dashed lines in Fig. \ref{fig:margR}). Nevertheless, there is a modest difference between the GR and MG best fit values, suggesting that the
impact of the interaction range $\lambda$ on the total likelihood
distribution is larger than for MACS 1206. Since the best fit is sensitive
to every small variation in the probability, the slight preference
of $\lambda\sim0.3$ (roughly three times the value from the best fit obtained in the previous case) is sufficient to move away the other parameters
from the GR values, but the excess is still not relevant enough to
affect the marginalized distributions
which remain in agreement with the analysis of Sartoris et al. (in prep.).

Following the same approach as in Sect. \ref{sec:noscreenM},
we combine the likelihood distribution generated by the \emph{MAMPOSSt}
method with the lensing posterior probability distribution $P(r_{200},r_{s})$
derived from the results of \cite{Umetsu16} and Caminha et al. (in prep). The blue contours and
the blue lines of Figure \ref{fig:margR} show what we obtain from
the analysis of $\mathcal{L}_{tot}$ defined in eq. \eqref{eq:l}.
No upper limits can be provided for $\lambda$ in this case. On the
contrary, the joint kinematic+lensing study indicates a preference
for large values of the effective interaction range, excluding at $\Delta\chi^{2}=2.71$
the region $\lambda\le0.14\,\mpc$, as we can see from the marginalized distribution
in the right bottom panel.

The origin of this behaviour is related to a slight tension
between the lensing and dynamics probability distributions in the
plane $(r_{s},r_{200})$.  Indeed, $P_{lens}(r_{200},r_{s})$ exhibits a
peak for $r_{200}=2.24\pm0.22\,\mpc,\,\,\,r_{s}=0.55\pm0.27\,\mpc$
(see the purple contours in the upper left panel of
Fig. \ref{fig:margR}), while the 2-dimensional dynamics distribution
has a maximum for $r_{s}=0.64\,\mpc$ and $r_{200}=2.65\,\mpc$. When combining the 2 analyses, we increase the
probability in the region of parameter space corresponding to
relatively large $\lambda$ values, while decreasing the probability for a
vanishing interaction range, as can be seen from the 2 dimensional distributions of Fig. \ref{fig:margR2}.

\begin{figure}
\includegraphics[width=1\textwidth]{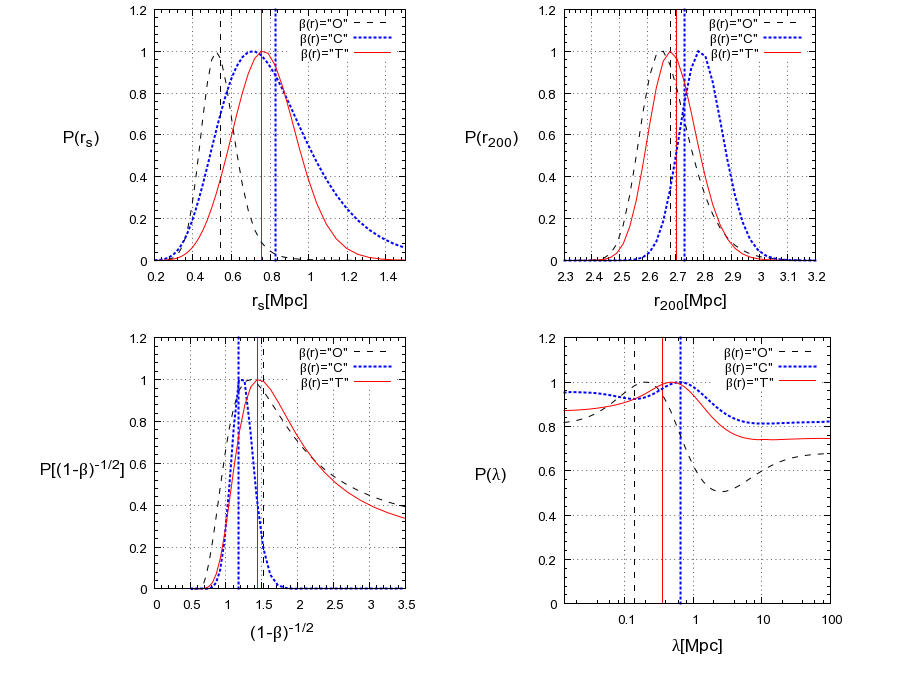}
\caption{\label{fig:sysR_beta}Effects on the marginalized likelihood distributions
of $r_{200}$, $r_{s}$, $\sigma_r/\sigma_{\theta}=1/\sqrt{1-\beta}$ and $\lambda$ from the dynamics analysis
of RXJ 2248 when changing the anisotropy model $\beta(r)$. Red solid curves: reference model \textquotedbl{}T\textquotedbl{}. Black dashed curves: \textquotedbl{}O\textquotedbl{} model. Blue dotted curves: \textquotedbl{}C\textquotedbl{} model.  The corresponding
vertical lines indicate the \emph{MAMPOSSt} best fit of each free
parameter. In the bottom left plot, $\beta\equiv\beta_{\infty}$ for the \textquotedbl{}T\textquotedbl{} and the \textquotedbl{}O\textquotedbl{} models, while $\beta\equiv\beta_C$ for the \textquotedbl{}C\textquotedbl{} model.}
\end{figure}

\begin{figure}
\includegraphics[width=1\textwidth]{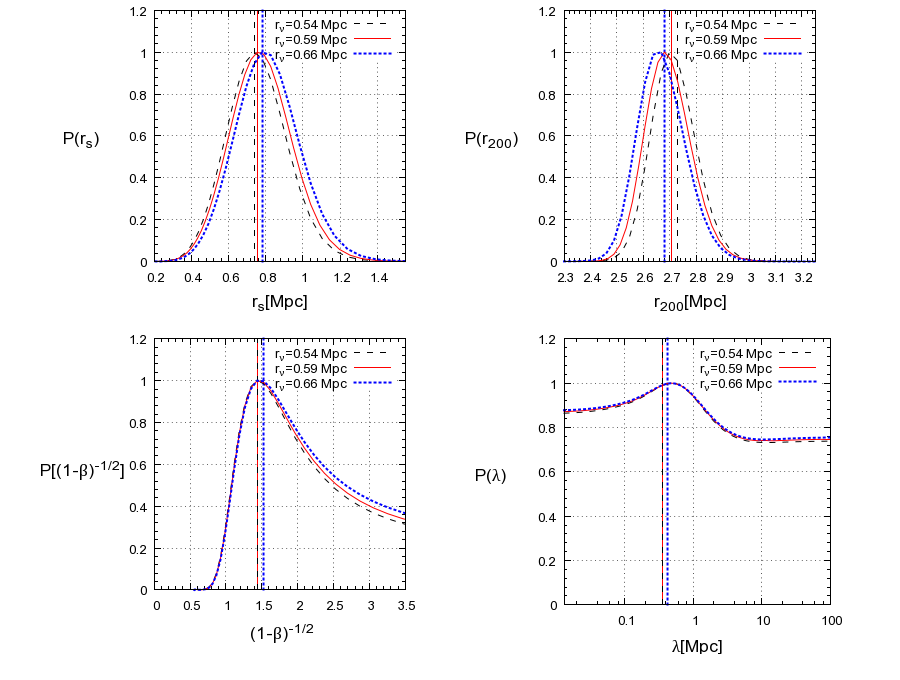}
\caption{\label{fig:sysR_rnu} Effects on the marginalized likelihood distributions
of $r_{200}$, $r_{s}$, $\sigma_r/\sigma_{\theta}=1/\sqrt{1-\beta}$ and $\lambda$ of the dynamics analysis 
of RXJ 2248 for the reference model \textquotedbl{}T\textquotedbl{} when changing the scale radius of the number density profile of the galaxies $r_{\nu}$ within the 68\% confidence region of the GR analysis. Red solid curves: $r_{\nu}$  fixed to the GR best fit value. Black dashed curves: $r_{\nu}$ fixed to
the lower limit of the 68\% confidence region. Blue dotted curves: $r_{\nu}$  settled to
the upper limit of the 68\% confidence region. The corresponding vertical
lines indicate the \emph{MAMPOSSt} best fit of each free parameter.}
\end{figure}

As before, we now study the effect of changing the scale radius of the
tracers $r_{\nu}$  within
the 68\% C.L. of the GR analysis and the anisotropy model $\beta(r)$. Fig.
\ref{fig:sysR_beta} shows the effect on the marginalized distributions due
to the different parametrization of the anisotropy profile while Fig. \ref{fig:sysR_rnu} displays the changes induced by $r_{\nu}$. The red curves in both Figures indicate the 
reference model (\textquotedbl{}T\textquotedbl{} in this case) with $r_{\nu}=0.59\,\mpc$. Again,
the largest effect on the results is obtained when changing the anisotropy model (Fig. \ref{fig:sysR_beta}). In this case both the position and the shape of the peak in the distribution of $\lambda$ are modified as a consequence of the variations in the $r_s$ marginalized likelihood; this is particularly evident for the model \textquotedbl{}O\textquotedbl{} which shows the largest discrepancy in the best fit value of $r_s$ compared to the reference model (vertical lines in the upper left plot of Fig. \ref{fig:sysR_beta}, see also Table \ref{Tab1}). Nevertheless, the tension with lensing results is not resolved when including these systematics. As shown in the five bottom rows of Table \ref{Tab1}, all the runs provide a lower limit on $\lambda$ which is larger than the reference model, except for the case of $r_{\nu}=0.66\,\mpc$ where the constraint on the interaction range relaxes from
$\lambda\ge0.30$ to $\lambda\ge0.06\,\mpc$ at $\Delta\chi^{2}=2.71$. 

Note that the results for $r_{\nu}=0.54\,\mpc$  is slightly preferred with respect to the case of $r_{\nu}=0.59\,\mpc$ (last row in Table \ref{Tab1}), similarly of what we found for MACS 1206.
In order to take into account the variation induced on the  NFW parameters $r_{s}$,
 $r_{200}$  by the changes in the anisotropy profile and in the scale radius of the number density profile, we consider the maximum difference 
between the best fit values of each run and of the reference model: 
\begin{equation}
r_{200}=[2.70_{-0.14}^{+0.10}(stat)\pm0.04(syst)]\,\mpc,\,\,\,\,\,\,\,\,\,\,\,\,\,r_{s}=[0.75_{-0.18}^{+0.14}\pm0.20(syst)]\,\mpc.
\end{equation}
As we did for the previous cluster, we estimate the systematic uncertainties in $\beta_{\infty}$ from the variation induced by $r_{\nu}$ only, obtaining:
\begin{equation}
\beta_{\infty}=1.43^{+0.59}_{-0.19}(stat)\pm0.10(syst). 
\end{equation}
\begin{table}[h]

\label{Tab1}
\begin{center}
\begin{tabular}{|cc|ccccc|}
   \hline
   {Anis.}& $r_{\nu}$&$r_{200}$&$r_{s}$&$(1-\beta_{C/\infty})^{-1/2}$&$\lambda$&$\Delta\chi^2$\\
    {model}      & [Mpc]  & [Mpc]  & [Mpc]&  & [Mpc]& \\
    \hline
   \hline
    &&&&{MACS 1206}&&\\
    \hline
    \hline 
   \rule[-4mm]{0mm}{1cm}
   $\bm{C}$&$\bm{0.63}$&$1.98\pm0.11$&$0.39^{+0.40}_{-0.16}$&$1.31^{+0.59}_{-0.19}$&$\le 1.61$&$0.0$\\
    \hline
    \rule[-4mm]{0mm}{1cm}
   $ O$&0.63&$1.93_{-0.12}^{+0.09}$&$0.30_{-0.07}^{+0.14}$&$1.51^{+0.69}_{-0.55}$&$\le1.31$&$0.72$\\
   \hline
    \rule[-4mm]{0mm}{1cm}
    $T$&0.63&$1.96_{-0.14}^{+0.07}$&$0.35_{-0.04}^{+0.33}$&$1.41^{+1.35}_{-0.29}$&$\le1.31$&$0.20$\\
    \hline
    \rule[-4mm]{0mm}{1cm}
  $ C$&0.74&$1.96_{-0.10}^{+0.11}$&$0.42_{-0.18}^{+0.42}$&$1.31_{-0.18}^{+0.59}$&$\le1.31$&$0.20$\\
   \hline
   \rule[-4mm]{0mm}{1cm}
  $ C$&0.54&$2.01^{+0.14}_{-0.07}$&$0.42_{-0.12}^{+0.28}$&$1.31_{-0.19}^{+0.59}$&$\le1.78$&$-0.14$\\
   \hline
   \hline
    &&&&{RXJ 2248}&&\\
    \hline
    \hline 
   \rule[-4mm]{0mm}{1cm}
     $\bm{T}$&$\bm{0.59}$&$2.70_{-0.14}^{+0.10}$&$0.75_{-0.18}^{+0.14}$&$1.43_{-0.24}^{+0.79}$&$\ge 0.14$&$0.0$\\
   \hline
    \rule[-4mm]{0mm}{1cm}
     $O$&$0.59$&$2.68_{-0.14}^{+0.10}$&$0.55_{-0.14}^{+0.11}$&$1.53_{-0.55}^{+0.71}$&$\ge 0.84$&$0.64$\\
   \hline
    \rule[-4mm]{0mm}{1cm}
     $C$&$0.59$&$2.74_{-0.06}^{+0.15}$&$0.83_{-0.32}^{+0.21}$&$1.18_{-0.14}^{+0.25}$&$\ge 0.80$&$0.44$\\
   \hline
    \rule[-4mm]{0mm}{1cm}
     $T$&0.66&$2.67\pm0.12$&$0.78_{-0.19}^{+0.26}$&$1.53_{-0.34}^{+0.69}$&$\ge0.06$&$0.20$\\
   \hline
   \rule[-4mm]{0mm}{1cm}
  $T$&0.54&$2.73^{+0.09}_{-0.15}$&$0.74_{-0.19}^{+0.24}$&$1.43_{-0.24}^{+0.66}$&$\ge0.19$&$-0.16$\\
   \hline
   \hline
\end{tabular}
\caption{Results on the free parameters of our analysis for the cluster MACS 1206 (first 5 rows) and RXJ 2248 (last 5 rows). The bold characters indicate the reference models adopted for each of the two clusters. The errors in $r_{200}$, $r_s$ and  $\sigma_r/\sigma_{\theta}=1/\sqrt{1-\beta}$ are the $68\%$ C.L. from the \emph{MAMPOSSt} procedure; the upper(lower) limits on $\lambda$ are obtained at $\Delta\chi^2=2.71$ from the joint kinematic+lensing analysis. The last column indicate the logarithmic difference between the likelihood of the model analysed and the likelihood of the reference model.}
\end{center}

\end{table}

\section{Discussion and conclusions}

\label{sec:discussion} In this work we have developed a procedure
to constrain modifications of gravity at the scales of galaxy clusters by
determining the time-time gravitational potential $\Phi(r)$ from
the analysis of the dynamics of the galaxies in the cluster. We have
focused on a particular sub-class of scalar-tensor theories, the $f(R)$
models, where the additional degree of freedom associated to the modification of gravity is expressed in terms of the interaction range $\lambda$; we have applied our method to the case of two dynamically relaxed clusters MACS 1206 at $z=0.44$ and RXJ 2248 at $z=0.35$, extensively analysed within the CLASH/CLASH-VLT collaborations. Assuming spherical symmetry, we have parametrized the cluster mass density profile as a NFW profile, characterized by
the scale radius $r_{s}$ and the virial radius $r_{200}$, constraining
the vector of parameters $(r_{s},r_{200},\lambda)$ with the modified
\emph{MAMPOSSt} code explained in Sect. \ref{sec:MAM}. Since in $f(R)$
gravity photons are not affected by the fifth force contribution we
have further combined our results with the information on the NFW
parameters $r_{s},r_{200}$ from the strong+weak lensing analysis
of ref. \cite{Umetsu16}. 

The results for the cluster MACS 1206 are in agreement with the GR
predictions, confirming the previous analysis of ref. \cite{Pizzuti16}.
When including lensing contribution, we obtain an upper limit of $[\lambda\le1.61(stat)+0.30(syst) \,\mpc$
at $\Delta\chi^{2}=2.71$; the bound takes into account variations
in the velocity anisotropy profile $\beta(r)$ and in the
scale radius of the galaxy number density profile $r_{\nu}$, which
enters in the kinematic determination of the gravitational potential
(or, equivalently, of the mass profile). 

\begin{figure}
\begin{center}
\includegraphics[width=0.5\textwidth]{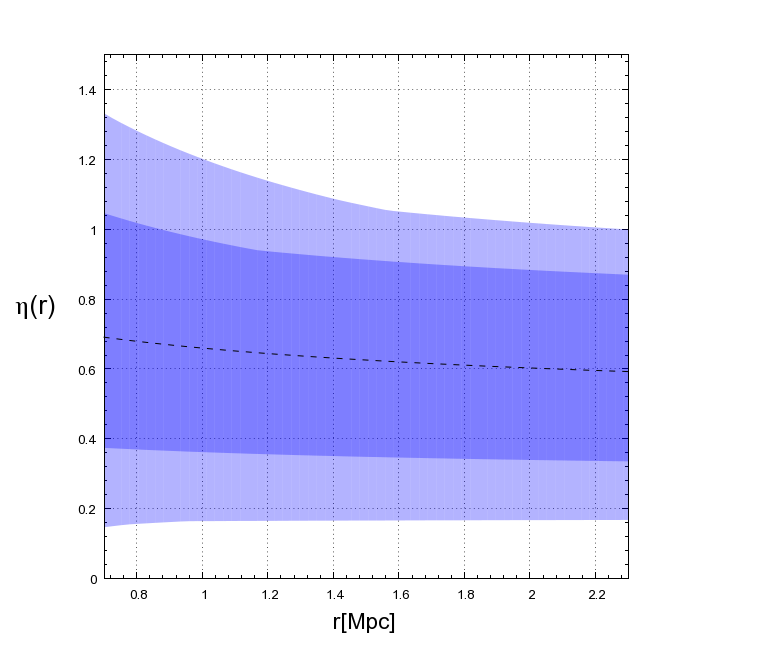} 
\end{center}
\caption{\label{fig:eta}Anisotropic stress $\eta$ in the radial range $[0.7\,\mpc-2.3\,\mpc]$ for the cluster RXJ 2248. The blue shaded region indicates the 68\% C.L. while the light blue area is the 90\% C.L. The best fit profile is given by the black dashed line.}
\end{figure}

A peculiar behaviour has instead been found in the case of RXJ 2248,
where the joint kinematic+lensing analysis shows a 2$\sigma$ preference
for values of $\lambda$ larger than 0. Including also systematics effects
due to changes in the anisotropy profile $\beta(r)$ and in $r_{\nu}$, we get a lower
bound $\lambda\ge0.12\,\mpc$ at $\Delta\chi^{2}=2.71$. This result arises
from a mild tension $< 1 \sigma$ between the dynamics and lensing determinations
of the mass profile in GR. Larger values of $r_{s}$ and $r_{200}$
are favoured by the dynamics analysis with respect to the lensing
results; the discrepancy could be interpreted as the additional contribution
of the fifth force which affects only the motion of non-relativistic
particles. 
In order to better investigate the effect of this tension on the Bardeen potential $\Phi$ and $\Psi$, we compute the anisotropic stress $\eta=\Psi/\Phi$ in the radial range covered by our analysis, following the approach of ref. \cite{Pizzuti16}. The results are shown in Fig. \ref{fig:eta}, where the blue and light blue areas represent the 68\% and the 90\%  confidence regions, respectively. As we can see, the discrepancy with the GR expected value $\eta=1$ is at more than 1$\sigma$ for $r\gtrsim1\,\mpc$. For $r=2.3\,\mpc$ we obtain $\eta=0.57\pm0.42$ at 90\% C.L. Interestingly, this is in agreement with the prediction of $f(R)$ models, in which the anisotropic stress is smaller than 1, reaching the value $\eta=1/2$ in the case of maximum deviations from GR.

The results from the analysis of RXJ 2248 point in the
opposite direction of what we have obtained for MACS 1206 (although the constraints we have derived on $\lambda$ from the 2 clusters are still compatible within the 90\% C.L.), highlighting
the need to check the systematics associated to the assumptions on which our method relies. The presence of possible substructures and departures from
spherical symmetry could in principle affect both the dynamics and
lensing analyses; nonetheless, we
stress again that both clusters belong to a sample of 20 X-rays selected
objects for their properties of apparent dynamical relaxation. 

It is important to point out that we have considered
only $f(R)$ models in which the screening mechanism works at scales
smaller (or much larger) than those investigated here, so that we could neglect the dependence of $m_{f_{R}}=1/\lambda$ on the local density.

In order to translate our results in a bound on the background scalaron
field $|\bar{f}_{,R}$|, we have to take into account how the chameleon
screening works once a particular model is fixed. As an example,
we focus on the Hu \& Sawicki model of ref. \cite{Hu2007}, in which
the functional form of $f(R)$ can be approximated by a power law:
\begin{equation}
f(R)\simeq6\Omega_{\Lambda}-\frac{f_{,R0}}{n}\frac{R_{0,b}^{n+1}}{R^{n}},\label{HSe}
\end{equation}
where we set $n=1$; $f_{,R0}<0$ is the background scalaron value
at present day. An accurate treatment of this situation requires a
full numerical solution of the nonlinear equation \eqref{eq:motion};
 we follow here a simple analytical approximation to describe
the screening mechanism in this case, just to show what we can obtain
with our method. We model the chameleon regime with an instantaneous
transition between the region of full screening $r\le S$, where $f_{,R}=0$,
and the linear region $r>S$, where the Newtonian potential is modified
according to eq. \eqref{phifr}. As explained in ref. \cite{Lomb12},
this can be achieved by taking $\delta f_{,R}=\min(\delta f_{,R}^{lin},|\bar{f}_{,R}|)$,
where $\delta f_{,R}^{lin}$ is the solution of the field equation \eqref{eq:motion}
with the linearization condition \eqref{eq:Rlin}.\\ 
In terms of $\lambda$, we define the effective interaction range constant and equal to the background value in the unscreened region $r>S$ and equal to zero for $r\le S$.

 We  run again the
\emph{MAMPOSSt} procedure computing the screening radius $S$ (which is given solving $\delta f_{,R}^{lin}(S)=\bar{f}_{,R}$ for each value of $r_s,\,r_{200},\lambda\equiv\bar{\lambda}_{f_R}$), and then requiring $\phi_{mg}(r)$ of eq. \eqref{eq:modmass} to be zero
  for $r\le S$. Combining the resulting likelihood with
the lensing posterior we get $\bar{\lambda}_{f_{R}}\le20\,\mpc$ and
$\bar{\lambda}_{f_{R}}\ge12\,\mpc$ at $\Delta\chi^{2}=2.71$ for
MACS 1206 and RXJ 2248, respectively. Now $\bar{\lambda}_{f_{R}}$
is related to the background field value through eq. \eqref{mass},
which is so constrained to be $|\bar{f}_{,R}(z=0.44)|\le4.0\times10^{-5}$
and $|\bar{f}_{,R}(z=0.35)|\ge1.4\times10^{-5}$. The first bound is
in agreement with current determinations of the magnitude of the background
scalaron, obtained using distance indicators at low-redshift (ref.
\mbox{%
\cite{Jain2013}%
}), galaxy cluster abundance (refs. %
\mbox{%
\cite{Schmidt09,Cataneo}%
}) and redshift space distortions (ref. %
\mbox{%
\cite{Xi15}%
}), which have tightened the upper limit up to $|\bar{f}_{,R0}|<10^{-6}$,
compatible with very small deviations from GR. On the other hand,
is not surprising that the effect of introducing screening for RXJ
2248 is to increase the tension we found, leading to a value of
$|\bar{f}_{,R0}|$ for this particular model which is totally inconsistent
with other constraints.
\begin{figure}
\begin{center}
\includegraphics[width=0.5\textwidth]{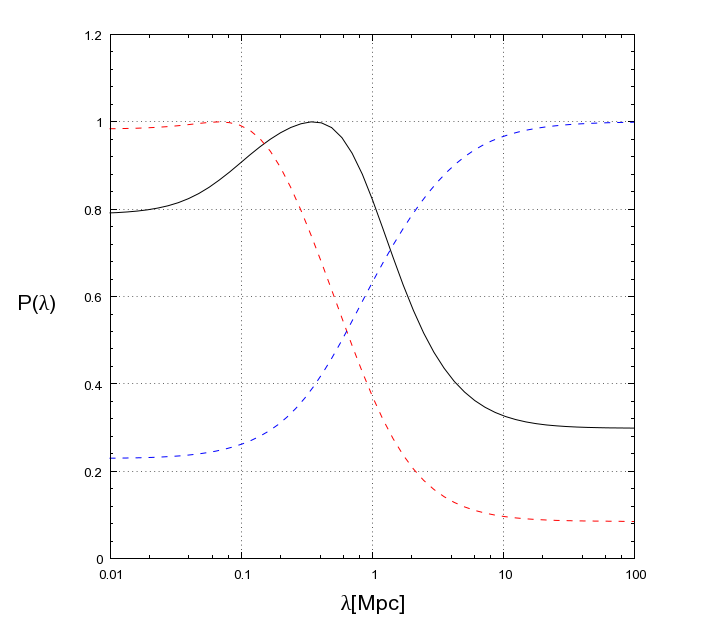} 
\end{center}
\caption{\label{fig:comb}Black solid line: combined likelihood obtained by multiplying the marginalized probability distributions from the analysis of the reference model for MACS1206 (red dashed line) and for RXJ 2248 (blue dashed line).}
\end{figure}

As the results we obtained are not in contradiction, we can in principle assume that the interaction range is constant in time between the redshifts of MACS 1206 and RXJ 2248 and combine the marginalized likelihood of $\lambda$ derived from the analysis of the reference model for each cluster. As we can see in Fig. \ref{fig:comb}, the apparent tension with GR is relaxed and we can still put an upper limit of $\lambda\le1.81\,\mpc$ at $\Delta\chi^2=1.0$.
In conclusion is clear that although two clusters
are obviously too few to produce stringent constraints, they are already
sufficient to show the potential of the method of combining dynamics
and lensing to test gravity. On the other hand, the tension we have
highlighted shows that it is necessary to model accurately the velocity
anisotropy and to take into account deviations from spherical symmetry
and from virialization before the method can be claimed to give a
robust determination of the anisotropic stress $\eta$ and its scale
dependence. This can be achieved by the analysis of simulated clusters, both in GR and in modified gravity, to better understand how the above mentioned effects influence our constraints. As a first step in this direction, in a subsequent work
we will estimate how many cluster similar to the one employed here
are necessary in order to constrain at the same time the profile parameters,
the velocity anisotropy, and the modified gravity parameters. 

Moreover, it is worth to notice that the analysis presented in this paper can be extended to all the generic scalar-tensor theories where the coupling constant $Q$ is not fixed and the lensing potential is still given by the Newtonian potential $(\Phi+\Psi)/2=\Phi_N$ (e.g. the Brans-Dicke k-essence (BDK) model of ref. \cite{BDK}), but the data we used here are not sufficient to get significant information on this kind of models.

Future imaging surveys, both from ground (e.g. LSST)
  and from space-borne telescopes (e.g. Euclid and WFIRST) will
  provide lensing mass reconstructions for thousands of clusters,
  although at a signal-to-noise level lower than that reached by the
  two clusters considered here. At the same time, the next generation
  of high-multiplexing spectrographs on 8m-class telescopes will allow
  a precise characterisation of the phase-space structure for a
  sizeable fraction of such clusters. This increase in statistics
  calls for the need of controlling the above mentioned systematics in
  the recovery of mass profiles from lensing and internal cluster
  dynamics, if we want to take full advantage of the their
  potentiality as powerful diagnostics for the nature of gravity on
  cosmological scales.

\medskip{}

\noindent \textbf{Acknowledgements.} 
We acknowledge financial support from the PRIN MIUR 2010-2011 (J91J12000450001)
grant, from the PRIN-MIUR 201278X4FL grant, from the PRIN-MIUR 2015W7KAWC\_002, from the ``InDark''
INFN Grant, from ``Consorzio per la Fisica di Trieste'', from the ``University of Trieste - Finanziamento di Ateneo per progetti di ricerca 
scientifica - FRA 2015" grant, from the
DFG TR33 ``The Dark Universe'' grant, and from PRIN-INAF 2014 1.05.01.94.02; K.U. acknowledges support from the Ministry of Science and Technology
of Taiwan under the grant MOST 103-2112-M-001-030-MY3.

 \bibliographystyle{JHEP}
\bibliography{master}

\providecommand{\href}[2]{#2}\begingroup\raggedright\begin{thebibliography}{10}

\bibitem{Reiss01}
A.~G. {Riess}, A.~V. {Filippenko}, P.~{Challis}, A.~{Clocchiatti},
  A.~{Diercks}, P.~M. {Garnavich}, R.~L. {Gilliland}, C.~J. {Hogan}, S.~{Jha},
  R.~P. {Kirshner}, B.~{Leibundgut}, M.~M. {Phillips}, D.~{Reiss}, B.~P.
  {Schmidt}, R.~A. {Schommer}, R.~C. {Smith}, J.~{Spyromilio}, C.~{Stubbs},
  N.~B. {Suntzeff}, and J.~{Tonry}, {\it {Observational Evidence from
  Supernovae for an Accelerating Universe and a Cosmological Constant}},  {\em
  \aj} {\bf 116} (Sept., 1998) 1009--1038,
  [\href{http://arxiv.org/abs/astro-ph/9805201}{{\tt astro-ph/9805201}}].

\bibitem{Perlmutter99}
S.~{Perlmutter}, G.~{Aldering}, G.~{Goldhaber}, R.~A. {Knop}, P.~{Nugent},
  P.~G. {Castro}, S.~{Deustua}, S.~{Fabbro}, A.~{Goobar}, D.~E. {Groom}, I.~M.
  {Hook}, A.~G. {Kim}, M.~Y. {Kim}, J.~C. {Lee}, N.~J. {Nunes}, R.~{Pain},
  C.~R. {Pennypacker}, R.~{Quimby}, C.~{Lidman}, R.~S. {Ellis}, M.~{Irwin},
  R.~G. {McMahon}, P.~{Ruiz-Lapuente}, N.~{Walton}, B.~{Schaefer}, B.~J.
  {Boyle}, A.~V. {Filippenko}, T.~{Matheson}, A.~S. {Fruchter}, N.~{Panagia},
  H.~J.~M. {Newberg}, W.~J. {Couch}, and T.~S.~C. {Project}, {\it {Measurements
  of {$\Omega$} and {$\Lambda$} from 42 High-Redshift Supernovae}},  {\em \apj}
  {\bf 517} (June, 1999) 565--586,
  [\href{http://arxiv.org/abs/astro-ph/9812133}{{\tt astro-ph/9812133}}].

\bibitem{Hu2007}
W.~{Hu} and I.~{Sawicki}, {\it {Models of f(R) cosmic acceleration that evade
  solar system tests}},  {\em \prd} {\bf 76} (Sept., 2007) 064004,
  [\href{http://arxiv.org/abs/0705.1158}{{\tt arXiv:0705.1158}}].

\bibitem{Dvali07}
G.~{Dvali}, S.~{Hofmann}, and J.~{Khoury}, {\it {Degravitation of the
  cosmological constant and graviton width}},  {\em \prd} {\bf 76} (Oct., 2007)
  084006, [\href{http://arxiv.org/abs/hep-th/0703027}{{\tt hep-th/0703027}}].

\bibitem{Lue01}
A.~{Lue}, R.~{Scoccimarro}, and G.~{Starkman}, {\it {Differentiating between
  modified gravity and dark energy}},  {\em \prd} {\bf 69} (Feb., 2004) 044005,
  [\href{http://arxiv.org/abs/astro-ph/0307034}{{\tt astro-ph/0307034}}].

\bibitem{Z09}
G.-B. {Zhao}, L.~{Pogosian}, A.~{Silvestri}, and J.~{Zylberberg}, {\it
  {Searching for modified growth patterns with tomographic surveys}},  {\em
  \prd} {\bf 79} (Apr., 2009) 083513,
  [\href{http://arxiv.org/abs/0809.3791}{{\tt arXiv:0809.3791}}].

\bibitem{H2013}
B.~{Hu}, M.~{Liguori}, N.~{Bartolo}, and S.~{Matarrese}, {\it {Parametrized
  modified gravity constraints after Planck}},  {\em \prd} {\bf 88} (Dec.,
  2013) 123514, [\href{http://arxiv.org/abs/1307.5276}{{\tt arXiv:1307.5276}}].

\bibitem{Za2010}
G.-B. {Zhao}, T.~{Giannantonio}, L.~{Pogosian}, A.~{Silvestri}, D.~J. {Bacon},
  K.~{Koyama}, R.~C. {Nichol}, and Y.-S. {Song}, {\it {Probing modifications of
  general relativity using current cosmological observations}},  {\em \prd}
  {\bf 81} (May, 2010) 103510, [\href{http://arxiv.org/abs/1003.0001}{{\tt
  arXiv:1003.0001}}].

\bibitem{Planckmod}
{Planck Collaboration}, P.~A.~R. {Ade}, N.~{Aghanim}, M.~{Arnaud},
  M.~{Ashdown}, J.~{Aumont}, C.~{Baccigalupi}, A.~J. {Banday}, R.~B.
  {Barreiro}, N.~{Bartolo}, and et~al., {\it {Planck 2015 results. XIV. Dark
  energy and modified gravity}},  {\em ArXiv e-prints} (Feb., 2015)
  [\href{http://arxiv.org/abs/1502.01590}{{\tt arXiv:1502.01590}}].

\bibitem{Ya2006}
K.~{Yamamoto}, B.~A. {Bassett}, R.~C. {Nichol}, Y.~{Suto}, and K.~{Yahata},
  {\it {Searching for modified gravity with baryon oscillations: From SDSS to
  wide field multiobject spectroscopy (WFMOS)}},  {\em \prd} {\bf 74} (Sept.,
  2006) 063525, [\href{http://arxiv.org/abs/astro-ph/0605278}{{\tt
  astro-ph/0605278}}].

\bibitem{J2012}
E.~{Jennings}, C.~M. {Baugh}, B.~{Li}, G.-B. {Zhao}, and K.~{Koyama}, {\it
  {Redshift-space distortions in f(R) gravity}},  {\em \mnras} {\bf 425}
  (Sept., 2012) 2128--2143, [\href{http://arxiv.org/abs/1205.2698}{{\tt
  arXiv:1205.2698}}].

\bibitem{Ferraro11}
S.~{Ferraro}, F.~{Schmidt}, and W.~{Hu}, {\it {Cluster abundance in f(R)
  gravity models}},  {\em \prd} {\bf 83} (Mar., 2011) 063503,
  [\href{http://arxiv.org/abs/1011.0992}{{\tt arXiv:1011.0992}}].

\bibitem{Pizzuti16}
L.~{Pizzuti}, B.~{Sartoris}, S.~{Borgani}, L.~{Amendola}, K.~{Umetsu},
  A.~{Biviano}, M.~{Girardi}, P.~{Rosati}, I.~{Balestra}, G.~B. {Caminha},
  B.~{Frye}, A.~{Koekemoer}, C.~{Grillo}, M.~{Lombardi}, A.~{Mercurio}, and
  M.~{Nonino}, {\it {CLASH-VLT: testing the nature of gravity with galaxy
  cluster mass profiles}},  {\em \jcap} {\bf 4} (Apr., 2016) 023,
  [\href{http://arxiv.org/abs/1602.03385}{{\tt arXiv:1602.03385}}].

\bibitem{Cataneo}
M.~{Cataneo}, D.~{Rapetti}, F.~{Schmidt}, A.~B. {Mantz}, S.~W. {Allen}, D.~E.
  {Applegate}, P.~L. {Kelly}, A.~{von der Linden}, and R.~G. {Morris}, {\it
  {New constraints on f (R ) gravity from clusters of galaxies}},  {\em \prd}
  {\bf 92} (Aug., 2015) 044009, [\href{http://arxiv.org/abs/1412.0133}{{\tt
  arXiv:1412.0133}}].

\bibitem{Dival2016}
E.~{Di Valentino}, A.~{Melchiorri}, and J.~{Silk}, {\it {Cosmological hints of
  modified gravity?}},  {\em \prd} {\bf 93} (Jan., 2016) 023513,
  [\href{http://arxiv.org/abs/1509.07501}{{\tt arXiv:1509.07501}}].

\bibitem{Will06}
C.~M. {Will}, {\it {The Confrontation between General Relativity and
  Experiment}},  {\em Living Reviews in Relativity} {\bf 9} (Mar., 2006)
  [\href{http://arxiv.org/abs/gr-qc/0510072}{{\tt gr-qc/0510072}}].

\bibitem{K2010}
J.~{Khoury}, {\it {Theories of Dark Energy with Screening Mechanisms}},  {\em
  ArXiv e-prints} (Nov., 2010) [\href{http://arxiv.org/abs/1011.5909}{{\tt
  arXiv:1011.5909}}].

\bibitem{Amendola2000}
L.~Amendola, {\it Coupled quintessence},  {\em Phys. Rev. D} {\bf 62} (Jul,
  2000) 043511.

\bibitem{Horn1974}
G.~W. Horndeski, {\it Second-order scalar-tensor field equations in a
  four-dimensional space},  {\em International Journal of Theoretical Physics}
  {\bf 10} (1974), no.~6 363--384.

\bibitem{K2004}
J.~{Khoury} and A.~{Weltman}, {\it {Chameleon cosmology}},  {\em \prd} {\bf 69}
  (Feb., 2004) 044026, [\href{http://arxiv.org/abs/astro-ph/0309411}{{\tt
  astro-ph/0309411}}].

\bibitem{Postman01}
M.~{Postman}, D.~{Coe}, N.~{Ben{\'{\i}}tez}, L.~{Bradley}, T.~{Broadhurst},
  M.~{Donahue}, H.~{Ford}, O.~{Graur}, G.~{Graves}, S.~{Jouvel},
  A.~{Koekemoer}, D.~{Lemze}, E.~{Medezinski}, A.~{Molino}, L.~{Moustakas},
  S.~{Ogaz}, A.~{Riess}, S.~{Rodney}, P.~{Rosati}, K.~{Umetsu}, W.~{Zheng},
  A.~{Zitrin}, M.~{Bartelmann}, R.~{Bouwens}, N.~{Czakon}, S.~{Golwala},
  O.~{Host}, L.~{Infante}, S.~{Jha}, Y.~{Jimenez-Teja}, D.~{Kelson},
  O.~{Lahav}, R.~{Lazkoz}, D.~{Maoz}, C.~{McCully}, P.~{Melchior},
  M.~{Meneghetti}, J.~{Merten}, J.~{Moustakas}, M.~{Nonino}, B.~{Patel},
  E.~{Reg{\"o}s}, J.~{Sayers}, S.~{Seitz}, and A.~{Van der Wel}, {\it {The
  Cluster Lensing and Supernova Survey with Hubble: An Overview}},  {\em \apjs}
  {\bf 199} (Apr., 2012) 25, [\href{http://arxiv.org/abs/1106.3328}{{\tt
  arXiv:1106.3328}}].

\bibitem{Rosati1}
P.~{Rosati}, I.~{Balestra}, C.~{Grillo}, A.~{Mercurio}, M.~{Nonino},
  A.~{Biviano}, M.~{Girardi}, E.~{Vanzella}, and {Clash-VLT Team}, {\it
  {CLASH-VLT: A VIMOS Large Programme to Map the Dark Matter Mass Distribution
  in Galaxy Clusters and Probe Distant Lensed Galaxies}},  {\em The Messenger}
  {\bf 158} (Dec., 2014) 48--53.

\bibitem{Mamon01}
G.~A. {Mamon}, A.~{Biviano}, and G.~{Bou{\'e}}, {\it {MAMPOSSt: Modelling
  Anisotropy and Mass Profiles of Observed Spherical Systems - I. Gaussian 3D
  velocities}},  {\em \mnras} {\bf 429} (Mar., 2013) 3079--3098,
  [\href{http://arxiv.org/abs/1212.1455}{{\tt arXiv:1212.1455}}].

\bibitem{Navarro}
J.~F. {Navarro}, C.~S. {Frenk}, and S.~D.~M. {White}, {\it {A Universal Density
  Profile from Hierarchical Clustering}},  {\em \apj} {\bf 490} (Dec., 1997)
  493--508, [\href{http://arxiv.org/abs/astro-ph/9611107}{{\tt
  astro-ph/9611107}}].

\bibitem{Li12}
B.~{Li}, G.-B. {Zhao}, and K.~{Koyama}, {\it {Haloes and voids in f(R)
  gravity}},  {\em \mnras} {\bf 421} (Apr., 2012) 3481--3487,
  [\href{http://arxiv.org/abs/1111.2602}{{\tt arXiv:1111.2602}}].

\bibitem{Umetsu16}
K.~{Umetsu}, A.~{Zitrin}, D.~{Gruen}, J.~{Merten}, M.~{Donahue}, and
  M.~{Postman}, {\it {CLASH: Joint Analysis of Strong-lensing, Weak-lensing
  Shear, and Magnification Data for 20 Galaxy Clusters}},  {\em \apj} {\bf 821}
  (Apr., 2016) 116, [\href{http://arxiv.org/abs/1507.04385}{{\tt
  arXiv:1507.04385}}].

\bibitem{Mukhanov01}
V.~{Mukhanov}, {\em {Physical Foundations of Cosmology}}.
\newblock Mar., 2001.

\bibitem{Bardeen01}
J.~M. {Bardeen}, {\it {Gauge-invariant cosmological perturbations}},  {\em
  \prd} {\bf 22} (Oct., 1980) 1882--1905.

\bibitem{Amendola2012}
L.~{Amendola}, M.~{Kunz}, M.~{Motta}, I.~D. {Saltas}, and I.~{Sawicki}, {\it
  {Observables and unobservables in dark energy cosmologies}},  {\em \prd} {\bf
  87} (Jan., 2013) 023501, [\href{http://arxiv.org/abs/1210.0439}{{\tt
  arXiv:1210.0439}}].

\bibitem{Defelice2011}
A.~{de Felice}, T.~{Kobayashi}, and S.~{Tsujikawa}, {\it {Effective
  gravitational couplings for cosmological perturbations in the most general
  scalar-tensor theories with second-order field equations}},  {\em Physics
  Letters B} {\bf 706} (Dec., 2011) 123--133,
  [\href{http://arxiv.org/abs/1108.4242}{{\tt arXiv:1108.4242}}].

\bibitem{navarro97}
J.~F. {Navarro}, C.~S. {Frenk}, and S.~D.~M. {White}, {\it {A Universal Density
  Profile from Hierarchical Clustering}},  {\em \apj} {\bf 490} (Dec., 1997)
  493--+, [\href{http://arxiv.org/abs/astro-ph/9}{{\tt astro-ph/9}}].

\bibitem{Buch01}
H.~A. {Buchdahl}, {\it {Non-linear Lagrangians and cosmological theory}},  {\em
  \mnras} {\bf 150} (1970) 1.

\bibitem{Cataneo16}
M.~{Cataneo}, D.~{Rapetti}, L.~{Lombriser}, and B.~{Li}, {\it {Cluster
  abundance in chameleon $f(R)$ gravity I: toward an accurate halo mass
  function prediction}},  {\em ArXiv e-prints} (July, 2016)
  [\href{http://arxiv.org/abs/1607.08788}{{\tt arXiv:1607.08788}}].

\bibitem{Pogosian10}
L.~{Pogosian} and A.~{Silvestri}, {\it {Erratum: Pattern of growth in viable
  f(R) cosmologies [Phys. Rev. D 77, 023503 (2008)]}},  {\em \prd} {\bf 81}
  (Feb., 2010) 049901.

\bibitem{Song2007}
Y.-S. {Song}, W.~{Hu}, and I.~{Sawicki}, {\it {Large scale structure of f(R)
  gravity}},  {\em \prd} {\bf 75} (Feb., 2007) 044004,
  [\href{http://arxiv.org/abs/astro-ph/0610532}{{\tt astro-ph/0610532}}].

\bibitem{Oyaizu08}
H.~{Oyaizu}, {\it {Nonlinear evolution of f(R) cosmologies. I. Methodology}},
  {\em \prd} {\bf 78} (Dec., 2008) 123523,
  [\href{http://arxiv.org/abs/0807.2449}{{\tt arXiv:0807.2449}}].

\bibitem{Schmidt10}
F.~{Schmidt}, {\it {Dynamical masses in modified gravity}},  {\em \prd} {\bf
  81} (May, 2010) 103002, [\href{http://arxiv.org/abs/1003.0409}{{\tt
  arXiv:1003.0409}}].

\bibitem{MamLok05}
G.~A. {Mamon} and E.~L. {{\L}okas}, {\it {Dark matter in elliptical galaxies -
  II. Estimating the mass within the virial radius}},  {\em \mnras} {\bf 363}
  (Nov., 2005) 705--722, [\href{http://arxiv.org/abs/astro-ph/0405491}{{\tt
  astro-ph/0405491}}].

\bibitem{Falco13}
M.~{Falco}, S.~H. {Hansen}, R.~{Wojtak}, and G.~A. {Mamon}, {\it {Why does the
  Jeans Swindle work?}},  {\em \mnras} {\bf 431} (Apr., 2013) L6--L9,
  [\href{http://arxiv.org/abs/1210.3363}{{\tt arXiv:1210.3363}}].

\bibitem{Biviano01}
A.~{Biviano}, P.~{Rosati}, I.~{Balestra}, A.~{Mercurio}, M.~{Girardi},
  M.~{Nonino}, C.~{Grillo}, M.~{Scodeggio}, D.~{Lemze}, D.~{Kelson},
  K.~{Umetsu}, M.~{Postman}, A.~{Zitrin}, O.~{Czoske}, S.~{Ettori}, A.~{Fritz},
  M.~{Lombardi}, C.~{Maier}, E.~{Medezinski}, S.~{Mei}, V.~{Presotto},
  V.~{Strazzullo}, P.~{Tozzi}, B.~{Ziegler}, M.~{Annunziatella},
  M.~{Bartelmann}, N.~{Benitez}, L.~{Bradley}, M.~{Brescia}, T.~{Broadhurst},
  D.~{Coe}, R.~{Demarco}, M.~{Donahue}, H.~{Ford}, R.~{Gobat}, G.~{Graves},
  A.~{Koekemoer}, U.~{Kuchner}, P.~{Melchior}, M.~{Meneghetti}, J.~{Merten},
  L.~{Moustakas}, E.~{Munari}, E.~{Reg{\H o}s}, B.~{Sartoris}, S.~{Seitz}, and
  W.~{Zheng}, {\it {CLASH-VLT: The mass, velocity-anisotropy, and
  pseudo-phase-space density profiles of the z = 0.44 galaxy cluster MACS
  J1206.2-0847}},  {\em \aap} {\bf 558} (Oct., 2013) A1,
  [\href{http://arxiv.org/abs/1307.5867}{{\tt arXiv:1307.5867}}].

\bibitem{girardi15}
M.~{Girardi}, A.~{Mercurio}, I.~{Balestra}, M.~{Nonino}, A.~{Biviano},
  C.~{Grillo}, P.~{Rosati}, M.~{Annunziatella}, R.~{Demarco}, A.~{Fritz},
  R.~{Gobat}, D.~{Lemze}, V.~{Presotto}, M.~{Scodeggio}, P.~{Tozzi},
  G.~{Bartosch Caminha}, M.~{Brescia}, D.~{Coe}, D.~{Kelson}, A.~{Koekemoer},
  M.~{Lombardi}, E.~{Medezinski}, M.~{Postman}, B.~{Sartoris}, K.~{Umetsu},
  A.~{Zitrin}, W.~{Boschin}, O.~{Czoske}, G.~{De Lucia}, U.~{Kuchner},
  C.~{Maier}, M.~{Meneghetti}, P.~{Monaco}, A.~{Monna}, E.~{Munari},
  S.~{Seitz}, M.~{Verdugo}, and B.~{Ziegler}, {\it {CLASH-VLT: Substructure in
  the galaxy cluster MACS J1206.2-0847 from kinematics of galaxy populations}},
   {\em \aap} {\bf 579} (July, 2015) A4,
  [\href{http://arxiv.org/abs/1503.05607}{{\tt arXiv:1503.05607}}].

\bibitem{Lemze01}
D.~{Lemze}, M.~{Postman}, S.~{Genel}, H.~C. {Ford}, I.~{Balestra},
  M.~{Donahue}, D.~{Kelson}, M.~{Nonino}, A.~{Mercurio}, A.~{Biviano},
  P.~{Rosati}, K.~{Umetsu}, D.~{Sand}, A.~{Koekemoer}, M.~{Meneghetti},
  P.~{Melchior}, A.~B. {Newman}, W.~A. {Bhatti}, G.~M. {Voit}, E.~{Medezinski},
  A.~{Zitrin}, W.~{Zheng}, T.~{Broadhurst}, M.~{Bartelmann}, N.~{Benitez},
  R.~{Bouwens}, L.~{Bradley}, D.~{Coe}, G.~{Graves}, C.~{Grillo}, L.~{Infante},
  Y.~{Jimenez-Teja}, S.~{Jouvel}, O.~{Lahav}, D.~{Maoz}, J.~{Merten},
  A.~{Molino}, J.~{Moustakas}, L.~{Moustakas}, S.~{Ogaz}, M.~{Scodeggio}, and
  S.~{Seitz}, {\it {The Contribution of Halos with Different Mass Ratios to the
  Overall Growth of Cluster-sized Halos}},  {\em \apj} {\bf 776} (Oct., 2013)
  91, [\href{http://arxiv.org/abs/1308.1675}{{\tt arXiv:1308.1675}}].

\bibitem{UmetsuMACS}
K.~{Umetsu}, E.~{Medezinski}, M.~{Nonino}, J.~{Merten}, A.~{Zitrin},
  A.~{Molino}, C.~{Grillo}, M.~{Carrasco}, M.~{Donahue}, A.~{Mahdavi},
  D.~{Coe}, M.~{Postman}, A.~{Koekemoer}, N.~{Czakon}, J.~{Sayers},
  T.~{Mroczkowski}, S.~{Golwala}, P.~M. {Koch}, K.-Y. {Lin}, S.~M. {Molnar},
  P.~{Rosati}, I.~{Balestra}, A.~{Mercurio}, M.~{Scodeggio}, A.~{Biviano},
  T.~{Anguita}, L.~{Infante}, G.~{Seidel}, I.~{Sendra}, S.~{Jouvel}, O.~{Host},
  D.~{Lemze}, T.~{Broadhurst}, M.~{Meneghetti}, L.~{Moustakas},
  M.~{Bartelmann}, N.~{Ben{\'{\i}}tez}, R.~{Bouwens}, L.~{Bradley}, H.~{Ford},
  Y.~{Jim{\'e}nez-Teja}, D.~{Kelson}, O.~{Lahav}, P.~{Melchior},
  J.~{Moustakas}, S.~{Ogaz}, S.~{Seitz}, and W.~{Zheng}, {\it {CLASH: Mass
  Distribution in and around MACS J1206.2-0847 from a Full Cluster Lensing
  Analysis}},  {\em \apj} {\bf 755} (Aug., 2012) 56,
  [\href{http://arxiv.org/abs/1204.3630}{{\tt arXiv:1204.3630}}].

\bibitem{Tiret01}
O.~{Tiret}, F.~{Combes}, G.~W. {Angus}, B.~{Famaey}, and H.~S. {Zhao}, {\it
  {Velocity dispersion around ellipticals in MOND}},  {\em \aap} {\bf 476}
  (Dec., 2007) L1--L4, [\href{http://arxiv.org/abs/0710.4070}{{\tt
  arXiv:0710.4070}}].

\bibitem{Mamon10}
G.~A. {Mamon}, A.~{Biviano}, and G.~{Murante}, {\it {The universal distribution
  of halo interlopers in projected phase space. Bias in galaxy cluster
  concentration and velocity anisotropy?}},  {\em \aap} {\bf 520} (Sept., 2010)
  A30, [\href{http://arxiv.org/abs/1003.0033}{{\tt arXiv:1003.0033}}].

\bibitem{Einasto65}
J.~{Einasto}, {\it {On the Construction of a Composite Model for the Galaxy and
  on the Determination of the System of Galactic Parameters}},  {\em Trudy
  Astrofizicheskogo Instituta Alma-Ata} {\bf 5} (1965) 87--100.

\bibitem{Hernquist01}
L.~{Hernquist}, {\it {An analytical model for spherical galaxies and bulges}},
  {\em \apj} {\bf 356} (June, 1990) 359--364.

\bibitem{Burkert01}
A.~{Burkert}, {\it {The Structure of Dark Matter Halos in Dwarf Galaxies}},
  {\em \apjl} {\bf 447} (July, 1995) L25,
  [\href{http://arxiv.org/abs/astro-ph/9504041}{{\tt astro-ph/9504041}}].

\bibitem{Geller99}
M.~J. {Geller}, A.~{Diaferio}, and M.~J. {Kurtz}, {\it {The Mass Profile of the
  Coma Galaxy Cluster}},  {\em \apjl} {\bf 517} (May, 1999) L23--L26,
  [\href{http://arxiv.org/abs/astro-ph/9903305}{{\tt astro-ph/9903305}}].

\bibitem{Zitirn01}
A.~{Zitrin}, T.~{Broadhurst}, R.~{Barkana}, Y.~{Rephaeli}, and
  N.~{Ben{\'{\i}}tez}, {\it {Strong-lensing analysis of a complete sample of 12
  MACS clusters at z > 0.5: mass models and Einstein radii}},  {\em \mnras}
  {\bf 410} (Jan., 2011) 1939--1956,
  [\href{http://arxiv.org/abs/1002.0521}{{\tt arXiv:1002.0521}}].

\bibitem{2015M}
J.~{Merten}, M.~{Meneghetti}, M.~{Postman}, K.~{Umetsu}, A.~{Zitrin},
  E.~{Medezinski}, M.~{Nonino}, A.~{Koekemoer}, P.~{Melchior}, D.~{Gruen},
  L.~A. {Moustakas}, M.~{Bartelmann}, O.~{Host}, M.~{Donahue}, D.~{Coe},
  A.~{Molino}, S.~{Jouvel}, A.~{Monna}, S.~{Seitz}, N.~{Czakon}, D.~{Lemze},
  J.~{Sayers}, I.~{Balestra}, P.~{Rosati}, N.~{Ben{\'{\i}}tez}, A.~{Biviano},
  R.~{Bouwens}, L.~{Bradley}, T.~{Broadhurst}, M.~{Carrasco}, H.~{Ford},
  C.~{Grillo}, L.~{Infante}, D.~{Kelson}, O.~{Lahav}, R.~{Massey},
  J.~{Moustakas}, E.~{Rasia}, J.~{Rhodes}, J.~{Vega}, and W.~{Zheng}, {\it
  {CLASH: The Concentration-Mass Relation of Galaxy Clusters}},  {\em \apj}
  {\bf 806} (June, 2015) 4, [\href{http://arxiv.org/abs/1404.1376}{{\tt
  arXiv:1404.1376}}].

\bibitem{Umetsu14}
K.~{Umetsu}, E.~{Medezinski}, M.~{Nonino}, J.~{Merten}, M.~{Postman},
  M.~{Meneghetti}, M.~{Donahue}, N.~{Czakon}, A.~{Molino}, S.~{Seitz},
  D.~{Gruen}, D.~{Lemze}, I.~{Balestra}, N.~{Ben{\'{\i}}tez}, A.~{Biviano},
  T.~{Broadhurst}, H.~{Ford}, C.~{Grillo}, A.~{Koekemoer}, P.~{Melchior},
  A.~{Mercurio}, J.~{Moustakas}, P.~{Rosati}, and A.~{Zitrin}, {\it {CLASH:
  Weak-lensing Shear-and-magnification Analysis of 20 Galaxy Clusters}},  {\em
  \apj} {\bf 795} (Nov., 2014) 163, [\href{http://arxiv.org/abs/1404.1375}{{\tt
  arXiv:1404.1375}}].

\bibitem{Zitrin2015}
A.~{Zitrin}, A.~{Fabris}, J.~{Merten}, P.~{Melchior}, M.~{Meneghetti},
  A.~{Koekemoer}, D.~{Coe}, M.~{Maturi}, M.~{Bartelmann}, M.~{Postman},
  K.~{Umetsu}, G.~{Seidel}, I.~{Sendra}, T.~{Broadhurst}, I.~{Balestra},
  A.~{Biviano}, C.~{Grillo}, A.~{Mercurio}, M.~{Nonino}, P.~{Rosati},
  L.~{Bradley}, M.~{Carrasco}, M.~{Donahue}, H.~{Ford}, B.~L. {Frye}, and
  J.~{Moustakas}, {\it {Hubble Space Telescope Combined Strong and Weak Lensing
  Analysis of the CLASH Sample: Mass and Magnification Models and Systematic
  Uncertainties}},  {\em \apj} {\bf 801} (Mar., 2015) 44,
  [\href{http://arxiv.org/abs/1411.1414}{{\tt arXiv:1411.1414}}].

\bibitem{Abell89}
G.~O. {Abell}, H.~G. {Corwin}, Jr., and R.~P. {Olowin}, {\it {A catalog of rich
  clusters of galaxies}},  {\em \apjs} {\bf 70} (May, 1989) 1--138.

\bibitem{Monna14}
A.~{Monna}, S.~{Seitz}, N.~{Greisel}, T.~{Eichner}, N.~{Drory}, M.~{Postman},
  A.~{Zitrin}, D.~{Coe}, A.~{Halkola}, S.~H. {Suyu}, C.~{Grillo}, P.~{Rosati},
  D.~{Lemze}, I.~{Balestra}, J.~{Snigula}, L.~{Bradley}, K.~{Umetsu},
  A.~{Koekemoer}, U.~{Kuchner}, L.~{Moustakas}, M.~{Bartelmann},
  N.~{Ben{\'{\i}}tez}, R.~{Bouwens}, T.~{Broadhurst}, M.~{Donahue}, H.~{Ford},
  O.~{Host}, L.~{Infante}, Y.~{Jimenez-Teja}, S.~{Jouvel}, D.~{Kelson},
  O.~{Lahav}, E.~{Medezinski}, P.~{Melchior}, M.~{Meneghetti}, J.~{Merten},
  A.~{Molino}, J.~{Moustakas}, M.~{Nonino}, and W.~{Zheng}, {\it {CLASH: z
  $\sim$ 6 young galaxy candidate quintuply lensed by the frontier field
  cluster RXC J2248.7-4431}},  {\em \mnras} {\bf 438} (Feb., 2014) 1417--1434,
  [\href{http://arxiv.org/abs/1308.6280}{{\tt arXiv:1308.6280}}].

\bibitem{Johnson14}
T.~L. {Johnson}, K.~{Sharon}, M.~B. {Bayliss}, M.~D. {Gladders}, D.~{Coe}, and
  H.~{Ebeling}, {\it {Lens Models and Magnification Maps of the Six Hubble
  Frontier Fields Clusters}},  {\em \apj} {\bf 797} (Dec., 2014) 48,
  [\href{http://arxiv.org/abs/1405.0222}{{\tt arXiv:1405.0222}}].

\bibitem{Richard14}
J.~{Richard}, M.~{Jauzac}, M.~{Limousin}, E.~{Jullo}, B.~{Cl{\'e}ment},
  H.~{Ebeling}, J.-P. {Kneib}, H.~{Atek}, P.~{Natarajan}, E.~{Egami},
  R.~{Livermore}, and R.~{Bower}, {\it {Mass and magnification maps for the
  Hubble Space Telescope Frontier Fields clusters: implications for
  high-redshift studies}},  {\em \mnras} {\bf 444} (Oct., 2014) 268--289,
  [\href{http://arxiv.org/abs/1405.3303}{{\tt arXiv:1405.3303}}].

\bibitem{Caminha16}
G.~B. {Caminha}, C.~{Grillo}, P.~{Rosati}, I.~{Balestra}, W.~{Karman},
  M.~{Lombardi}, A.~{Mercurio}, M.~{Nonino}, P.~{Tozzi}, A.~{Zitrin},
  A.~{Biviano}, M.~{Girardi}, A.~M. {Koekemoer}, P.~{Melchior},
  M.~{Meneghetti}, E.~{Munari}, S.~H. {Suyu}, K.~{Umetsu}, M.~{Annunziatella},
  S.~{Borgani}, T.~{Broadhurst}, K.~I. {Caputi}, D.~{Coe},
  C.~{Delgado-Correal}, S.~{Ettori}, A.~{Fritz}, B.~{Frye}, R.~{Gobat},
  C.~{Maier}, A.~{Monna}, M.~{Postman}, B.~{Sartoris}, S.~{Seitz},
  E.~{Vanzella}, and B.~{Ziegler}, {\it {CLASH-VLT: A highly precise strong
  lensing model of the galaxy cluster RXC J2248.7-4431 (Abell S1063) and
  prospects for cosmography}},  {\em \aap} {\bf 587} (Mar., 2016) A80,
  [\href{http://arxiv.org/abs/1512.04555}{{\tt arXiv:1512.04555}}].

\bibitem{Gruen13}
D.~{Gruen}, F.~{Brimioulle}, S.~{Seitz}, C.-H. {Lee}, J.~{Young},
  J.~{Koppenhoefer}, T.~{Eichner}, A.~{Riffeser}, V.~{Vikram}, T.~{Weidinger},
  and A.~{Zenteno}, {\it {Weak lensing analysis of RXC J2248.7-4431}},  {\em
  \mnras} {\bf 432} (June, 2013) 1455--1467,
  [\href{http://arxiv.org/abs/1304.0764}{{\tt arXiv:1304.0764}}].

\bibitem{Merten15}
J.~{Merten}, M.~{Meneghetti}, M.~{Postman}, K.~{Umetsu}, A.~{Zitrin},
  E.~{Medezinski}, M.~{Nonino}, A.~{Koekemoer}, P.~{Melchior}, D.~{Gruen},
  L.~A. {Moustakas}, M.~{Bartelmann}, O.~{Host}, M.~{Donahue}, D.~{Coe},
  A.~{Molino}, S.~{Jouvel}, A.~{Monna}, S.~{Seitz}, N.~{Czakon}, D.~{Lemze},
  J.~{Sayers}, I.~{Balestra}, P.~{Rosati}, N.~{Ben{\'{\i}}tez}, A.~{Biviano},
  R.~{Bouwens}, L.~{Bradley}, T.~{Broadhurst}, M.~{Carrasco}, H.~{Ford},
  C.~{Grillo}, L.~{Infante}, D.~{Kelson}, O.~{Lahav}, R.~{Massey},
  J.~{Moustakas}, E.~{Rasia}, J.~{Rhodes}, J.~{Vega}, and W.~{Zheng}, {\it
  {CLASH: The Concentration-Mass Relation of Galaxy Clusters}},  {\em \apj}
  {\bf 806} (June, 2015) 4, [\href{http://arxiv.org/abs/1404.1376}{{\tt
  arXiv:1404.1376}}].

\bibitem{Melchior15}
P.~{Melchior}, E.~{Suchyta}, E.~{Huff}, M.~{Hirsch}, T.~{Kacprzak},
  E.~{Rykoff}, D.~{Gruen}, R.~{Armstrong}, D.~{Bacon}, K.~{Bechtol}, G.~M.
  {Bernstein}, S.~{Bridle}, J.~{Clampitt}, K.~{Honscheid}, B.~{Jain},
  S.~{Jouvel}, E.~{Krause}, H.~{Lin}, N.~{MacCrann}, K.~{Patton}, A.~{Plazas},
  B.~{Rowe}, V.~{Vikram}, H.~{Wilcox}, J.~{Young}, J.~{Zuntz}, T.~{Abbott},
  F.~B. {Abdalla}, S.~S. {Allam}, M.~{Banerji}, J.~P. {Bernstein}, R.~A.
  {Bernstein}, E.~{Bertin}, E.~{Buckley-Geer}, D.~L. {Burke}, F.~J.
  {Castander}, L.~N. {da Costa}, C.~E. {Cunha}, D.~L. {Depoy}, S.~{Desai},
  H.~T. {Diehl}, P.~{Doel}, J.~{Estrada}, A.~E. {Evrard}, A.~F. {Neto},
  E.~{Fernandez}, D.~A. {Finley}, B.~{Flaugher}, J.~A. {Frieman},
  E.~{Gaztanaga}, D.~{Gerdes}, R.~A. {Gruendl}, G.~R. {Gutierrez}, M.~{Jarvis},
  I.~{Karliner}, S.~{Kent}, K.~{Kuehn}, N.~{Kuropatkin}, O.~{Lahav}, M.~A.~G.
  {Maia}, M.~{Makler}, J.~{Marriner}, J.~L. {Marshall}, K.~W. {Merritt}, C.~J.
  {Miller}, R.~{Miquel}, J.~{Mohr}, E.~{Neilsen}, R.~C. {Nichol}, B.~D. {Nord},
  K.~{Reil}, N.~A. {Roe}, A.~{Roodman}, M.~{Sako}, E.~{Sanchez}, B.~X.
  {Santiago}, R.~{Schindler}, M.~{Schubnell}, I.~{Sevilla-Noarbe},
  E.~{Sheldon}, C.~{Smith}, M.~{Soares-Santos}, M.~E.~C. {Swanson}, A.~J.
  {Sypniewski}, G.~{Tarle}, J.~{Thaler}, D.~{Thomas}, D.~L. {Tucker},
  A.~{Walker}, R.~{Wechsler}, J.~{Weller}, and W.~{Wester}, {\it {Mass and
  galaxy distributions of four massive galaxy clusters from Dark Energy Survey
  Science Verification data}},  {\em \mnras} {\bf 449} (May, 2015) 2219--2238,
  [\href{http://arxiv.org/abs/1405.4285}{{\tt arXiv:1405.4285}}].

\bibitem{Gomez12}
P.~L. {G{\'o}mez}, L.~E. {Valkonen}, A.~K. {Romer}, E.~{Lloyd-Davies},
  T.~{Verdugo}, C.~M. {Cantalupo}, M.~D. {Daub}, J.~H. {Goldstein}, C.~L.
  {Kuo}, A.~E. {Lange}, M.~{Lueker}, W.~L. {Holzapfel}, J.~B. {Peterson},
  J.~{Ruhl}, M.~C. {Runyan}, C.~L. {Reichardt}, and K.~{Sabirli}, {\it {Optical
  and X-Ray Observations of the Merging Cluster AS1063}},  {\em \aj} {\bf 144}
  (Sept., 2012) 79.

\bibitem{Biviano06}
A.~{Biviano} and P.~{Salucci}, {\it {The radial profiles of the different mass
  components in galaxy clusters}},  {\em \aap} {\bf 452} (June, 2006) 75--81,
  [\href{http://arxiv.org/abs/astro-ph/0511309}{{\tt astro-ph/0511309}}].

\bibitem{Bartelmann96}
M.~{Bartelmann}, {\it {Arcs from a universal dark-matter halo profile.}},  {\em
  \aap} {\bf 313} (Sept., 1996) 697--702,
  [\href{http://arxiv.org/abs/astro-ph/9602053}{{\tt astro-ph/9602053}}].

\bibitem{Beers90}
T.~C. {Beers}, K.~{Flynn}, and K.~{Gebhardt}, {\it {Measures of location and
  scale for velocities in clusters of galaxies - A robust approach}},  {\em
  \aj} {\bf 100} (July, 1990) 32--46.

\bibitem{Lomb12}
L.~{Lombriser}, K.~{Koyama}, G.-B. {Zhao}, and B.~{Li}, {\it {Chameleon f(R)
  gravity in the virialized cluster}},  {\em \prd} {\bf 85} (June, 2012)
  124054, [\href{http://arxiv.org/abs/1203.5125}{{\tt arXiv:1203.5125}}].

\bibitem{Jain2013}
B.~{Jain}, V.~{Vikram}, and J.~{Sakstein}, {\it {Astrophysical Tests of
  Modified Gravity: Constraints from Distance Indicators in the Nearby
  Universe}},  {\em \apj} {\bf 779} (Dec., 2013) 39,
  [\href{http://arxiv.org/abs/1204.6044}{{\tt arXiv:1204.6044}}].

\bibitem{Schmidt09}
F.~{Schmidt}, M.~{Lima}, H.~{Oyaizu}, and W.~{Hu}, {\it {Nonlinear evolution of
  f(R) cosmologies. III. Halo statistics}},  {\em \prd} {\bf 79} (Apr., 2009)
  083518--+, [\href{http://arxiv.org/abs/0812.0545}{{\tt arXiv:0812.0545}}].

\bibitem{Xi15}
L.~Xu, {\it Constraints on $f(r)$ gravity through the redshift space
  distortion},  {\em Phys. Rev. D} {\bf 91} (Mar, 2015) 063008.

\bibitem{BDK}
H.~{Kim}, {\it {Brans-Dicke theory as a unified model for dark matter-dark
  energy}},  {\em \mnras} {\bf 364} (Dec., 2005) 813--822,
  [\href{http://arxiv.org/abs/astro-ph/0408577}{{\tt astro-ph/0408577}}].

\end{thebibliography}\endgroup

\end{document}